\documentclass[a4paper,11pt]{article}
\usepackage{jheppub} % for details on the use of the package, please see the JINST-author-manual
\usepackage{lineno}

\title{\boldmath Unstable Instantons in $A$-model Localization}

\author{Emil Hakan Leeb-Lundberg}
\affiliation{Department of Physics, Boğaziçi University \\
Istanbul, Turkey}
\affiliation{Niels Bohr Institute, Copenhagen University\\
Blegdamsvej 17, Copenhagen, 2100, Denmark}

\emailAdd{emilleeblundberg@gmail.com}

\abstract{We apply localization techniques to $A$-twisted $\mathcal{N}=(2,2)$
theories of vector multiplets on $S^{2}$. We derive formulae for
$A$-model partition functions and correlators as integrals
along a real contour, as opposed to a complex one. Using the correlator
formula, we successfully recover the unstable instanton partition
function of pure two-dimensional Yang-Mills theory from the vacuum
expectation value of an $A$-model operator. }

\begin{document}
\maketitle
\flushbottom

\section{Introduction} \label{sec:1}

Supersymmetric theories provide a mathematically tractable framework
to investigate subtle phenomena exhibited by quantum field theories
found in nature. Instantons, for instance, mediate tunneling in quantum
mechanics and anomalies in quantum chromodynamics. They also appear
in supersymmetric gauge theories on compact manifolds, where supersymmetric
localization \cite{Pestun:2007rz} permits the exact evaluation of
BPS observables, including the partition function. A valuable feature
of supersymmetric localization is that it admits the introduction
of exact deformations, leading to different mathematical descriptions
of the same observable. The advantage of having multiple descriptions
is that evaluating an observable may be straightforward in one description
and challenging or impossible in another.

In this paper, we apply localization techniques to the $A$-model
 on $S^{2}$, specifically $\mathcal{N}=(2,2)$ gauge theories with
a vector-like R-symmetry, topologically $A$-twisted on $S^{2}$.
This was one of the subjects studied by Benini and Zaffaroni in \cite{Benini:2015noa,Benini:2016hjo}
and Closset, Cremonesi, and Park in \cite{Closset:2015rna}, building
on earlier localization computations of twisted and un-twisted $\mathcal{N}=(2,2)$
gauge theories on $S^{2}$ \cite{Benini:2012ui,Doroud:2012xw,Doroud:2013pka,Gomis:2012wy,Melnikov:2006kb}.
The authors of \cite{Benini:2015noa,Benini:2016hjo,Closset:2015rna} used 
localization techniques to derive a contour integral
formula for correlators of the $A$-model  on a closed two-dimensional
genus $g$ Riemannian manifold $\Sigma_{g}$. However, there is at
least one $A$-model  observable that remains challenging to
evaluate using this contour integral formula. This observable is the
unstable instanton partition function of two-dimensional pure Yang-Mills
($\text{YM}_{2}$) theory on $\Sigma_{g}$. We derive a formula in
which $A$-model  correlators are described as integrals along
a real contour, rather than a complex contour, then recover the $\text{YM}_{2}$
partition function (\ref{eq:Winv9}) from the $A$-model  in a straightforward
manner. 

The results of \cite{Benini:2015noa,Benini:2016hjo,Closset:2015rna}
include contour integral formulae for partition functions and correlators
of $A$-twisted and $\Omega$-deformed theories on compact manifolds
in two, three, and four dimensions. These have proven useful for $\text{AdS}_{4}$
black hole microstate counting \cite{Benini:2015eyy} and checking
dualities \cite{Closset:2017vvl}. In addition to modern studies,
there are classic examples of $A$-model  localization in the
context of cohomological theories on $\Sigma_{g}$. The examples primarily
concern the $A$-model  vector multiplet in Wess-Zumino gauge,
which is related to the standard multiplet of two-dimensional cohomological
gauge theory by field redefinitions. In the seminal paper \cite{Witten:1992xu},
Witten introduced non-abelian localization and described its application
to path integrals of cohomological theories of the standard multiplet
on $\Sigma_{g}$. He further established a map between cohomological
gauge theory and $\text{YM}_{2}$ theory by relating the vacuum expectation
value (vev) of a standard multiplet operator to the $\text{YM}_{2}$
partition function. The standard multiplet has also appeared in recent
examples of $A$-model  localization, and this perspective
has proven useful in studying the moduli space of the BPS vortex on
$\Sigma_{g}$ \cite{Ohta:2019odi}.

The purpose of this paper is to derive formulae for partition functions
and correlators of the $A$-model  vector multiplet on $S^{2}$
using localization techniques established in \cite{Witten:1992xu},
then use the correlator formula to recover the $\text{YM}_{2}$ partition
function from the vev of an $A$-model  operator. Our localization
scheme differs from the one in \cite{Benini:2015noa,Benini:2016hjo,Closset:2015rna}
by an exact term, and leads to a different mathematical representation
of the same $A$-model  formulae. In particular, we provide
evidence that partition functions and correlators of the $A$-model
 can be regarded as integrals along a real contour. A key feature
of our derivation is that the localization techniques and resulting
formulae are, in principle, straightforward. Our description of $A$-model
 correlators proves advantageous in evaluating the vev of the $A$-model
 operator that yields the $\text{YM}_{2}$ partition function. Although
performing the same computation with the contour integral formula
for $A$-model correlators is conceivable, it is more challenging,
both conceptually and technically. Essentially, the derivation is
the $A$-model  version of non-abelian localization of cohomological
gauge theory on $S^{2}$ and incorporates the map to $\text{YM}_{2}$
theory. 

More generally, this work aims to address a discrepancy among distinct
supersymmetric localization techniques and contribute to their potential
reconciliation. The discrepancy lies between: \emph{i)} conventional
supersymmetric localization techniques established in \cite{Pestun:2007rz},
\emph{ii)} non-abelian localization techniques established in \cite{Witten:1992xu},
and \emph{iii)} supersymmetric localization techniques employing JK
residues established in \cite{Benini:2013nda}. The authors of \cite{Benini:2015noa,Benini:2016hjo,Closset:2015rna}
demonstrated that conventional localization techniques omit the necessary
path integral contribution of unstable instantons in the $A$-model
 on $S^{2}$, and that these contributions could be retained with
localization techniques employing JK residues. This paper demonstrates
that non-abelian localization techniques provide an alternative method
to retain the necessary contributions of unstable instantons to the
path integral of the $A$-model  on $S^{2}$. While the applicability
of non-abelian localization techniques to the $A$-model  vector
multiplet on $S^{2}$ is recognized, the scope of each of the three
approaches to supersymmetric localization remains less clear.

The formula for correlators of the $A$-model  vector multiplet
on $S^{2}$ is of the schematic form \cite{Benini:2015noa,Benini:2016hjo,Closset:2015rna}
\begin{equation}
\left\langle \mathcal{O}\right\rangle =\frac{1}{\left|W\right|}\sum_{\mathfrak{m}\in\Lambda_{\text{coch}}}\,\int_{\mathcal{C}}\text{d}u\,\mathcal{O}(u)\,Z_{\text{class}}(u;\mathfrak{m})\,Z_{\text{1-loop}}(u;\mathfrak{m}).\label{eq:intro_AmodelCorrelator}
\end{equation}
Here, $\mathfrak{m}\in\Lambda_{\text{coch}}$ is the GNO quantized
gauge flux on $S^{2}$, $\Lambda_{\text{coch}}$ is the cocharacter/GNO
lattice in the Cartan subalgebra $\mathfrak{h}$ of $\mathfrak{g}=\text{Lie}(G)$,
$u\in\mathfrak{h}_{\mathbb{C}}$ is a continuous complex modulus that
parametrizes the bosonic scalar $\sigma$ in the vector multiplet,
$\mathfrak{h}_{\mathbb{C}}$ is the complexification of $\mathfrak{h}$,
$\left|W\right|$ is the order of the Weyl group, $\mathcal{O}$ is
a gauge-invariant operator, $Z_{\text{class}}$ is the classical contribution,
and $Z_{\text{1-loop}}$ is the one-loop contribution. For a reference,
see equation 1.13 in \cite{Closset:2015rna}, where the integration
contour and classical contribution have already been specified. 

As explained in section 3 of \cite{Benini:2016hjo} and section 4
of \cite{Closset:2015rna}, the integral in (\ref{eq:intro_AmodelCorrelator})
is evaluated by integrating $u$ along the complex Jeffrey-Kirwan
(JK) contour $\mathcal{C}=\mathcal{C}_{\text{JK}}$, resulting in
a sum over JK residues. For abelian gauge groups, evaluating the sums
over gauge fluxes and JK residues yields the final answer. For non-abelian
gauge groups, additional steps are necessary to account for all contributions
\cite{Benini:2016hjo}. This involves first summing over the gauge
fluxes and then omitting solutions of the Bethe Ansatz Equations (BAE)
for which the one-loop determinant is zero. The result of this procedure
is a sum over a subset of BAE solutions, whose evaluation yields the
final answer.

Although our $A$-model  correlator formula is schematically
of the form (\ref{eq:intro_AmodelCorrelator}), it differs in a few
crucial aspects. Most importantly, our $u$ is a continuous real modulus
that can be integrated along the real contour $\mathcal{C}=\mathbb{R}$.
Furthermore, our $\mathfrak{m}$ labels BPS Yang-Mills connections
that are zero-action configurations of the localizing term. We derive
(\ref{eq:intro_AmodelCorrelator}) in a manner that differs from the
original derivation at several stages in the localization computation.
Aside from the unconventional choice of localizing term prescribed
by \cite{Witten:1992xu}, we employ standard localization techniques
established in \cite{Pestun:2007rz,Kapustin:2009kz,Benini:2012ui}.
We evaluate the one-loop determinant in (\ref{eq:intro_AmodelCorrelator})
mode-by-mode using monopole spherical harmonics on the $S^{2}$. 

Our recovery of the $\text{YM}_{2}$ partition function from the vev
of an $A$-model  operator extends previous derivations in
a number of ways. In \cite{Witten:1992xu}, the $\text{YM}_{2}$ partition
function was recovered from the vev of an operator of the standard
multiplet for the rank-one gauge groups $G=SU(2)$ and $G=SO(3)$.
In \cite{Blau:1993hj}, Blau and Thompson recovered the $\text{YM}_{2}$
partition function from the vev of an operator of the basic multiplet
for the higher-rank gauge group $G=SU(N)$, where the basic multiplet
is a part of the standard multiplet. We recover the $\text{YM}_{2}$
partition function from the vev of an operator of the $A$-model
 vector multiplet for non-simply connected versions of $G=SU(N)$.
In particular, our derivation extends the results of \cite{Witten:1992xu,Blau:1993hj}
to include higher-rank non-simply connected versions of the gauge
group and incorporates a phase factor that was absent in the one-loop
determinant. 

\subsection{Overview}

This paper is organized as follows. In section 2, we briefly review
the $\mathcal{N}=(2,2)$ supersymmetric vector multiplet on $\mathbb{R}^{2}$,
the topological $A$-twist, and the $A$-model  vector multiplet
on $S^{2}$. We then proceed to $A$-model  localization, addressing
the construction of localizing terms, the moduli space, gauge-fixing,
and the locus expansion. After deriving the gauge-fixed localizing
term to quadratic order in fluctuations, we evaluate one-loop determinants
using monopole spherical harmonics and conclude with the integral
formulae for partition functions and correlators of the $A$-model
 vector multiplet. In section 3, we use the $A$-model  correlator
formula to recover the two-dimensional Yang-Mills partition function
from the vev of an $A$-model  operator. We detail our choices
for the operator insertion and classical contribution, evaluate the
sum and integral, eliminate the Weyl group invariance, and use representation
theory formulae to recover the final partition function. In section
4, we discuss our results and propose future directions.
        
%\section{$A$-model  Localization} \label{sec:2}
\section[\textit{A}-model Localization]{\textit{A}-model Localization} \label{sec:2}

\paragraph{Conventions}

Our conventions are similar to those of \cite{Benini:2012ui,Benini:2016qnm}.
The Euclidean flat-space metric is $\delta_{\mu\nu}$ for $\mu,\nu=1,2$.
We use complex coordinates $z=x^{1}+ix^{2}$ , $\bar{z}=x^{1}-ix^{2}$
, where $\delta_{z\bar{z}}=\frac{1}{2}$, $\delta_{zz}=\delta_{\bar{z}\bar{z}}=0$,
holomorphic vectors are $X_{z}=\frac{1}{2}(X_{1}-iX_{2}$), anti-holomorphic
vectors are $X_{\bar{z}}=\frac{1}{2}(X_{1}+iX_{2})$, and $X_{\mu}X^{\mu}=2(X_{z}X_{\bar{z}}+X_{\bar{z}}X_{z})$. 

An anti-commuting Dirac spinor with a lower index denotes a ``column vector'' 
\begin{equation}
\lambda_{\alpha}=\left({\lambda_{+}\atop \lambda_{-}}\right),    
\end{equation}
where $\lambda_{+}$ and $\lambda_{-}$ are Weyl spinors of charge
$+\frac{1}{2}$ and $-\frac{1}{2}$, respectively, under $\text{Spin}(2)\simeq\text{U}(1)_{E}$.
Spinor indices are raised and lowered as 
\begin{equation}
\lambda^{\alpha}=C^{\alpha\beta}\lambda_{\beta}\,,\quad\lambda_{\alpha}=C_{\alpha\beta}\lambda^{\beta}\,,\quad\implies\quad\lambda^{+}=\lambda_{-}\;,\quad\lambda^{-}=-\lambda_{+}\;.
\end{equation}
where $C^{\alpha\beta}$ and $C_{\alpha\beta}$ are anti-symmetric
tensors and $C_{+-}=-C^{-+}=1$. Spinors are contracted
as 
\begin{equation}
\psi\lambda\equiv\psi^{\alpha}\lambda_{\alpha}\;,\qquad\quad\psi\gamma^{\mu}\lambda\equiv\psi^{\alpha}(\gamma^{\mu})_{\alpha}\,^{\beta}\lambda_{\beta}\;,
\end{equation}
and the gamma matrices are 
\begin{equation}
(\gamma^{1})_{\alpha}\,^{\beta}=\begin{pmatrix}0 & 1\\
1 & 0
\end{pmatrix}\;,\qquad(\gamma^{2})_{\alpha}\,^{\beta}=\begin{pmatrix}0 & -i\\
i & 0
\end{pmatrix}\;,\qquad(\gamma^{3})_{\alpha}\,^{\beta}=\begin{pmatrix}1 & 0\\
0 & -1
\end{pmatrix}\;,
\end{equation}
where $\gamma^{3}=-i\gamma^{1}\gamma^{2}$. 

We use the physics convention where all gauge fields are hermitian.
The gauge covariant derivative acting on fields in the adjoint representation
and the field strength are defined as
\begin{equation}
D_{\mu}=\partial_{\mu}-i[A_{\mu},\cdot],\qquad F_{\mu\nu}=\partial_{\mu}A_{\nu}-\partial_{\nu}A_{\mu}-i[A_{\mu},A_{\nu}].
\end{equation}
Together, these satisfy $F_{\mu\nu}=i[D_{\mu},D_{\nu}]$. The relationship
to the math convention is $A_{\mu}=iA_{\mu}^{\prime}$ and $F_{\mu\nu}=iF_{\mu\nu}^{\prime}$
where $A_{\mu}^{\prime}$ and $F_{\mu\nu}^{\prime}$ are anti-hermitian.
A gauge transformation with parameter $\vartheta$ reads 
\begin{equation}
\delta_{\vartheta}=i\left[\vartheta,\cdot\right],\qquad\delta_{\vartheta}A_{\mu}=D_{\mu}\vartheta .
\end{equation}

\subsection{$\mathcal{N}=(2,2)$ supersymmetry on $\mathbb{R}^{2}$}

The two-dimensional $\mathcal{N}=(2,2)$ vector multiplet in Wess-Zumino
gauge has 8 real components. It comprises of a connection $A_{\mu}$,
the real scalars $\sigma_{1}$ and $\sigma_{2}$, a Dirac spinor $\lambda$,
and a real auxiliary scalar $D$, all in the adjoint representation
of the Lie algebra $\mathfrak{g}$. In Euclidean signature, the fields
are complexified. The real scalars $\sigma_{1}$ and $\sigma_{2}$
sit in two generically complex scalars $\sigma$ and $\tilde{\sigma}$
, and we have a Dirac spinor $\tilde{\lambda}$ that is independent
of $\lambda$. 

The four generators of $\mathcal{N}=(2,2)$ supersymmetry are combined
as 
\begin{equation}
\delta=\frac{1}{\sqrt{2}}(\epsilon^{\alpha}Q_{\alpha}+\tilde{\epsilon}^{\alpha}\widetilde{Q}_{\alpha})=\frac{1}{\sqrt{2}}(\epsilon_{+}Q_{-}-\epsilon_{-}Q_{+}+\tilde{\epsilon}_{+}\widetilde{Q}_{-}-\tilde{\epsilon}_{-}\widetilde{Q}_{+}),
\end{equation}
where $\epsilon_{\pm},\tilde{\epsilon}_{\pm}$ are commuting, and
$Q_{\pm}$,$\widetilde{Q}_{\pm},$ are anti-commuting. The supersymmetry
variations of the vector multiplet are 
\begin{equation}
\begin{array}{ccl}
\delta A_{\mu} & = & \frac{i}{\sqrt{2}}(\epsilon\gamma_{\mu}\tilde{\lambda}+\widetilde{\epsilon}\gamma_{\mu}\lambda) ,\\
\delta\sigma & = & -\sqrt{2}\epsilon P_{-}\tilde{\lambda}+\sqrt{2}\widetilde{\epsilon}P_{+}\lambda ,\\
\delta\tilde{\sigma} & = & -\sqrt{2}\epsilon P_{+}\tilde{\lambda}+\sqrt{2}\tilde{\epsilon}P_{-}\lambda ,\\
\delta D & = & -\frac{i}{\sqrt{2}}D_{\mu}\left(\epsilon\gamma^{\mu}\tilde{\lambda}-\widetilde{\epsilon}\gamma^{\mu}\lambda\right)-\frac{i}{\sqrt{2}}\epsilon\left(\begin{array}{cc}
[\sigma,\cdot] & 0\\
0 & [\tilde{\sigma},\cdot]
\end{array}\right)\tilde{\lambda}-\frac{i}{\sqrt{2}}\widetilde{\epsilon}\left(\begin{array}{cc}
[\tilde{\sigma},\cdot] & 0\\
0 & [\sigma,\cdot]
\end{array}\right)\lambda ,\\
\delta\lambda & = & \frac{i}{\sqrt{2}}\left(-iD+\left(-F_{12}+\frac{1}{2}[\sigma,\tilde{\sigma}]\right)\gamma_{3}+D_{\mu}\left(\begin{array}{cc}
\sigma & 0\\
0 & \tilde{\sigma}
\end{array}\right)\gamma^{\mu}\right)\epsilon ,\\
\delta\tilde{\lambda} & = & -\frac{i}{\sqrt{2}}\left(-iD+\left(F_{12}+\frac{1}{2}[\sigma,\tilde{\sigma}]\right)\gamma_{3}+D_{\mu}\left(\begin{array}{cc}
\tilde{\sigma} & 0\\
0 & \sigma
\end{array}\right)\gamma^{\mu}\right)\widetilde{\epsilon} ,\\
\delta F_{12} & = & -\frac{1}{\sqrt{2}}D_{\mu}(\epsilon\gamma^{\mu}\gamma_{3}\tilde{\lambda}+\widetilde{\epsilon}\gamma^{\mu}\gamma_{3}\lambda),
\end{array}\label{eq:susyR2}
\end{equation}
where $P_{\pm}=\frac{1}{2}(1-\gamma_{3})$. The supersymmetry algebra
is 
\begin{equation}
\left\{ Q_{\alpha},\widetilde{Q}_{\beta}\right\} =\left(2\gamma^{\mu}P_{\mu}\right)_{\alpha\beta},\quad\left\{ \delta_{\epsilon},\delta_{\tilde{\epsilon}}\right\} =i\mathcal{L}_{\epsilon\gamma^{\mu}\tilde{\epsilon}},
\end{equation}
where $P_{\mu}=-i\partial_{\mu}$, $\mathcal{L}_{\epsilon\gamma^{\mu}\tilde{\epsilon}}$
is the Lie derivative along $\epsilon\gamma^{\mu}\tilde{\epsilon}$,
and the central charges have been set to zero. 

Under the vector-like R-symmetry group $\text{U}(1)_{V}$, the supercharges $(Q_{+},Q_{-},\tilde{Q}_{+},\tilde{Q}_{-})$
have charge $(-1,-1,1,1)$, the parameters $(\epsilon_{+},\epsilon_{-},\tilde{\epsilon}_{+},\tilde{\epsilon}_{-})$
have charge $(1,1,-1,-1)$, and the fields $(A_{\mu},\sigma,\tilde{\sigma},D,\lambda_{+},\lambda_{-},\tilde{\lambda}_{+},\tilde{\lambda}_{-})$
have charge $(0,0,0,0,1,1,-1,-1)$. 

%The charges under the
%vector-like R-symmetry are 

%\begin{table}[htb]
%\centering{}%
%\begin{tabular}{c|cc|cc|cccccc}
% & $\epsilon_{\pm}$ & $\tilde{\epsilon}_{\pm}$ & $Q_{\pm}$ & $\widetilde{Q}_{\pm}$ & %$A_{\mu}$ & $\sigma$ & $\tilde{\sigma}$ & $\lambda_{\pm}$ & $\tilde{\lambda}_{\pm}$ & %$D$\tabularnewline
%\hline 
%$U(1)_{V}$ & 1 & -1 & -1 & 1 & 0 & 0 & 0 & 1 & -1 & 0\tabularnewline
%\end{tabular}
%\end{table}

\subsection{$A$-model  on $S^{2}$}

The $A$-model  on $S^{2}$ is defined by performing the topological
$A$-twist of the $\mathcal{N}=\left(2,2\right)$ supersymmetry on $\mathbb{R}^{2}$
\cite{Witten:1988xj,Closset:2014pda,Benini:2016qnm}. The metric,
vielbein, and spin connection of the $S^{2}$ are 
\begin{equation}
\text{d}s^{2}=R^{2}\left(\text{d}\theta^{2}+\sin^{2}\theta\,\text{d}\varphi^{2}\right),\quad e^{1}=R\text{d}\theta,\quad e^{2}=R\sin\theta\text{d}\varphi,\quad\omega=\cos\theta\text{d}\varphi ,
\end{equation}
where $R$ is the radius of the $S^{2}$, $\varphi$ is $2\pi$ periodic,
and $0\leq\theta\leq\pi$. We further define the holomorphic vielbein
$e^{z}=e^{1}+ie^{2}$, and denote the volume form on the $S^{2}$
by $\text{d}\mu$. The $A$-twist is a solution of the Killing spinor
equations of the $S^{2}$ that preserves half of the $\mathcal{N}=\left(2,2\right)$
supercharges of $\mathbb{R}^{2}$. The solution is \cite{Benini:2016qnm}
\begin{equation}
A_{\mu}^{V}=\frac{1}{2}\omega_{\mu},\qquad\epsilon=\begin{pmatrix}0\\
\epsilon_{-}
\end{pmatrix},\qquad\tilde{\epsilon}=\begin{pmatrix}\tilde{\epsilon}_{+}\\
0
\end{pmatrix},\qquad\mathcal{H}=0,\qquad\widetilde{\mathcal{H}}=0 ,
\end{equation}
for constant $\epsilon_{-}$ and $\tilde{\epsilon}_{+}$, i.e., $\partial_{\mu}\epsilon_{-}=0$
and $\partial_{\mu}\tilde{\epsilon}_{+}=0$. Here, $A_{\mu}^{V}$
is a connection of a background $U(1)_{V}$ symmetry, $\omega_{\mu}$
is the spin connection on $S^{2}$, and $\mathcal{H},\widetilde{\mathcal{H}}$
are bosonic scalars.

Under the $A$-twisted Lorentz group, the supercharges $(Q_{+},Q_{-},\tilde{Q}_{+},\tilde{Q}_{-})$ have charge $(0,-2,2,0)$, the supersymmetry parameters $(\epsilon_{+},\epsilon_{-},\tilde{\epsilon}_{+},\tilde{\epsilon}_{-})$ have charge $(2,0,0,-2)$,  and the fields $(A_{z},A_{\bar{z}},\sigma,\tilde{\sigma},D,\lambda_{+},\lambda_{-},\tilde{\lambda}_{+},\tilde{\lambda}_{-})$  have charge $(2,-2,0,0,0,2,0,0,-2)$.  Quantities with charge $0$ are scalars, those with charge $+2$ are holomorphic vectors, and those with charge $-2$ are anti-holomorphic vectors. In particular, after the $A$-twist, $Q_{+},\widetilde{Q}_{-},\lambda_{-},\tilde{\lambda}_{+}$ are scalars, $\lambda_{+},\widetilde{Q}_{+}$ are  holomorphic vectors, and $\tilde{\lambda}_{-},Q_{-}$ are anti-holomorphic vectors. To make the new spins in the  $A$-model  manifest, we relabel the supercharges and fermions as 
\begin{equation}
Q:=Q_{+},\qquad\widetilde{Q}:=\widetilde{Q}_{-},\qquad\lambda_{z}:=\lambda_{+},\qquad\lambda:=\lambda_{-},\qquad\tilde{\lambda}:=\tilde{\lambda}_{+},\qquad\lambda_{\bar{z}}:=\tilde{\lambda}_{-}.
\end{equation}

The $A$-model  vector multiplet on $S^{2}$ consists of a
connection $A_{\mu}$, the complex scalars $\sigma$ and $\tilde{\sigma}$,
a fermionic vector $\lambda_{\mu}$, the fermionic scalars $\lambda$
and $\tilde{\lambda}$, and an auxiliary field $D$. The $A$-model
 scalar supercharge 
\begin{equation}
Q_{A}=Q+\widetilde{Q}
\end{equation}
acts on the fields of the vector multiplet as
\begin{equation}
\begin{array}{l}
Q_{A}A_{z}=i\lambda_{z},\\
Q_{A}A_{\bar{z}}=i\lambda_{\bar{z}},\\
Q_{A}\sigma=0,\\
Q_{A}\tilde{\sigma}=2\lambda-2\tilde{\lambda},\\
Q_{A}D=-2iD_{z}\lambda_{\bar{z}}+2iD_{\bar{z}}\lambda_{z}-i[\sigma,\lambda+\tilde{\lambda}],\\
Q_{A}\lambda_{z}=-2iD_{z}\sigma,\\
Q_{A}\lambda=-D-i\star F+\frac{i}{2}\left[\sigma,\tilde{\sigma}\right],\\
Q_{A}\tilde{\lambda}=-D-i\star F-\frac{i}{2}\left[\sigma,\tilde{\sigma}\right],\\
Q_{A}\lambda_{\bar{z}}=-2iD_{\bar{z}}\sigma,\\
Q_{A}\star F=2D_{z}\lambda_{\bar{z}}-2D_{\bar{z}}\lambda_{z},
\end{array}\label{eq:QAtransf}
\end{equation}
where the scalar $\star F=F_{12}=-2iF_{z\bar{z}}$ is the Hodge dual
of the field strength. The transformations (\ref{eq:QAtransf}) are
obtained by setting $\epsilon_{+}=\tilde{\epsilon}_{-}=0$ and $\epsilon_{-}=\tilde{\epsilon}_{+}=1$
in (\ref{eq:susyR2}), and relabeling the fields. See appendix \ref{appendix} for details.

We define $\tilde{D}=D+i\star F$, such that transformations of $D,\lambda,\tilde{\lambda}$
under $Q_{A}$ are modified as 
\begin{equation}
\begin{array}{lcc}
Q_{A}\tilde{D} & = & -i[\sigma,\lambda+\tilde{\lambda}] ,\\
Q_{A}\lambda & = & -\tilde{D}+\frac{i}{2}[\sigma,\tilde{\sigma}],\\
Q_{A}\tilde{\lambda} & = & -\tilde{D}-\frac{i}{2}[\sigma,\tilde{\sigma}].
\end{array}
\end{equation}
The algebra of the $A$-model  scalar supercharge is
\begin{equation}
\begin{array}{ccl}
Q_{A}^{2}A_{\mu} & = & 2D_{\mu}\sigma ,\\
Q_{A}^{2}\sigma & = & 0 ,\\
Q_{A}^{2}\Phi & = & 2i\left[\sigma,\Phi\right],
\end{array}
\end{equation}
where $\Phi$ denotes $\tilde{\sigma},\tilde{D},\lambda_{\mu},\lambda,\tilde{\lambda}$.
Since $Q_{A}$ squares to a gauge transformation with parameter $\vartheta=2\sigma$,
it is nilpotent when acting on gauge-invariant functionals of the
fields of the vector multiplet.

The $A$-model  actions are similar to the standard supersymmetric
actions of the $\mathcal{N}=(2,2)$ vector multiplet on $\mathbb{R}^{2}$
with the fields relabeled, but the exact terms differ. In $\mathcal{N}=(2,2)$
supersymmetry, there is the D-term Yang-Mills action $S_{\text{SYM}}$,
the F-term twisted chiral superpotential action $S_{\widetilde{W}}$,
and its conjugate $S_{\overline{\widetilde{W}}}$. $S_{\text{SYM}}$
is exact under $Q_{\pm}$ and $\widetilde{Q}_{\pm}$, $S_{\widetilde{W}}$
is exact under $Q_{-}$ and $\widetilde{Q}_{+}$, and $S_{\overline{\widetilde{W}}}$
is exact under $Q_{+}$ and $\widetilde{Q}_{-}$. In the $A$-model
, $S_{\text{SYM}}$, $S_{\widetilde{W}}$, and $S_{\overline{\widetilde{W}}}$
are $Q_{A}$-closed, but only $S_{\text{SYM}}$ and $S_{\overline{\widetilde{W}}}$
are $Q_{A}$-exact. 

The standard supersymmetric actions of the $A$-model  vector
multiplet are the Yang-Mills action

\begin{equation}
\begin{array}{lcl}
S_{\text{SYM}} & = & \int\text{d}\mu\,\text{Tr}\,\Bigg(\frac{1}{2}\left(\star F\right)^{2}+\frac{1}{2}\left(D\right)^{2}+\frac{1}{2}D_{\mu}\tilde{\sigma}D^{\mu}\sigma+\frac{1}{8}\left[\sigma,\tilde{\sigma}\right]^{2}\\
 &  & \qquad\qquad\qquad+2i\lambda_{z}D_{\bar{z}}\tilde{\lambda}-2i\lambda_{\bar{z}}D_{z}\lambda+i\lambda_{\bar{z}}\left[\tilde{\sigma},\lambda_{z}\right]-i\tilde{\lambda}[\sigma,\lambda]\Bigg),
\end{array}\label{eq:Ssym}
\end{equation}
and the twisted chiral superpotential actions 
\begin{equation}
\begin{array}{lcl}
S_{\widetilde{W}} & = & \int\mathrm{d}\mu\,\text{Tr}\,\left(\frac{\partial\widetilde{W}}{\partial\sigma_{a}}(-i\star F-D)^{a}-\frac{1}{2}\frac{\partial^{2}\widetilde{W}}{\partial\sigma_{a}\partial\sigma_{b}}\lambda_{\bar{z}}^{a}\lambda_{z}^{b}\right),\end{array}\label{eq:SWa}
\end{equation}

\begin{equation}
\begin{array}{lcl}
S_{\overline{\widetilde{W}}} & = & \int\mathrm{d}\mu\,\text{Tr}\,\left(\frac{\partial\overline{\widetilde{W}}}{\partial\tilde{\sigma}_{a}}(-i\star F+D)^{a}+\frac{1}{2}\frac{\partial^{2}\overline{\widetilde{W}}}{\partial\tilde{\sigma}_{a}\partial\tilde{\sigma}_{b}}\lambda^{a}\tilde{\lambda}^{b}\right).\end{array}\label{eq:SWb}
\end{equation}
Here $a,b$ are adjoint indices, $\widetilde{W}$ is a holomorphic
functional of $\sigma$, and $\overline{\widetilde{W}}$ is a anti-holomorphic
functional of $\tilde{\sigma}$. The supersymmetric Yang-Mills action
(\ref{eq:Ssym}) serves as the localizing term in \cite{Benini:2015noa,Benini:2016hjo,Closset:2015rna}. 

To relate the $A$-model  vector multiplet to the standard
multiplet of two-dimensional cohomological gauge theory on $S^{2}$,
the fields are redefined as 
\begin{equation}
\begin{array}{lcccl}
A_{\mu}=ia_{\mu}, &  & \tilde{\sigma}=\tilde{\phi}, &  & \lambda=-\frac{\eta}{4}-\chi,\\
\lambda_{\mu}=-\psi_{\mu}, &  & \tilde{D}=iH, &  & \tilde{\lambda}=\frac{\eta}{4}-\chi.\\
\sigma=-\frac{\phi}{2},
\end{array}\label{eq:dictStaVec}
\end{equation}
After this field redefinition, the transformations (\ref{eq:QAtransf})
become those of the standard multiplet 
\begin{equation}
\begin{array}{lcccl}
\delta A_{\mu}=i\epsilon\psi_{\mu}, &  & \delta\tilde{\phi}=i\epsilon\eta, &  & \delta\chi=\epsilon H,\\
\delta\psi_{\mu}=-\epsilon D_{\mu}\phi, &  & \delta\eta=\epsilon[\phi,\tilde{\phi}], &  & \delta H=i\epsilon\left[\phi,\chi\right],\\
\delta\phi=0,
\end{array}\label{eq:staplet}
\end{equation}
where $\delta=-i\,\epsilon\,Q_{A}$ and we have reinstated $\epsilon=\epsilon_{-}=\tilde{\epsilon}_{+}$. 

Two actions play a central role in the original non-abelian localization
of cohomological gauge theories. These are the $\delta$-exact localizing
term and the $\delta$-closed physical Yang-Mills action in equations
3.27 and 3.43 of \cite{Witten:1992xu}, respectively. To describe
the localizing term, we first define the fermionic functional
\begin{equation}
\begin{array}{ccl}
V(t) & = & V+tV^{\prime}\\
 & = & \int\mathrm{d}\mu\,\text{Tr}\,\left(\frac{1}{2}\chi\left(H-2\star F\right)+D_{\mu}\tilde{\phi}\psi^{\mu}\right)-t\int\mathrm{d}\mu\,\text{Tr}\,\chi\lambda
\end{array}\label{eq:V=000026Vprime}
\end{equation}
where $t$ is a parameter. Acting with $\delta$ on $V(t)$, we obtain
the localizing term
\begin{equation}
\begin{array}{lcl}
S_{\text{coh}}(t) & = & \delta V(t)\\
 & = & \int\mathrm{d}\mu\,\text{Tr}\,\bigg(\frac{1}{2}\left(H-\star F-\tilde{\phi}t\right)^{2}-\frac{1}{2}\left(\star F+t\tilde{\phi}\right)^{2}-D_{\mu}\tilde{\phi}D^{\mu}\phi\\
 &  & \qquad\qquad\qquad+\frac{i}{2}\chi\left[\phi,\chi\right]-i\chi\star D\psi+iD_{\mu}\eta\psi^{\mu}+i\left[\psi_{\mu},\tilde{\phi}\right]\psi^{\mu}-it\chi\eta\bigg)
\end{array}\label{eq:ScohLoc}
\end{equation}
The physical Yang-Mills action is
\begin{equation}
S_{\text{YM}}=-\underset{\omega}{\underbrace{\frac{1}{4\pi^{2}}\int\mathrm{Tr}\left(i\phi F+\frac{1}{2}\psi\wedge\psi\right)}}-\underset{\Theta}{\varepsilon\;\underbrace{\frac{1}{8\pi^{2}}\int\mathrm{d}\mu\,\text{Tr}\,\phi^{2}}},\label{eq:ScohYM}
\end{equation}
where $\omega$ and $\Theta$ are BRST-invariant operators and $\varepsilon$
is a real positive parameter. Although (\ref{eq:ScohYM}) is a functional
of the fields of the standard cohomological multiplet, it is equivalent
to two-dimensional Yang-Mills theory. 

Comparing the localizing term (\ref{eq:ScohLoc}) to the $Q_{A}$-exact
$A$-model  actions (\ref{eq:Ssym}) and (\ref{eq:SWb}), we
see that $S_{\text{coh}}(t)$ is nearly the same as $S_{\text{SYM}}+tS_{\overline{\widetilde{W}}}$
with quadratic $\overline{\widetilde{W}}(\tilde{\sigma})=\text{Tr}\tilde{\sigma}^{2}$.
However, $S_{\text{SYM}}$ includes a bosonic scalar commutator term
that is not present in $S_{\text{coh}}(t)$. Comparing the physical
Yang-Mills action (\ref{eq:ScohYM}) to the $Q_{A}$-closed $A$-model
 superpotential (\ref{eq:SWa}), we see that the $\omega$ term in
$S_{\text{YM}}$ is nearly the same as $S_{\widetilde{W}}$ with quadratic
$\widetilde{W}(\sigma)=\text{Tr}\sigma^{2}$. However, $S_{\text{YM}}$
includes $\Theta$, while $S_{\widetilde{W}}$ does not. We will use
these comparisons to construct the $A$-model  vector multiplet
analogs of (\ref{eq:ScohLoc}) and (\ref{eq:ScohYM}). 

\subsection{Localizing term}

To construct the localizing term, we define a generic fermionic functional
with parameters, then act on it with the localizing supercharge $Q_{A}$.
For specific values of the parameters, the generic localizing term
reduces to the localizing terms prescribed by either \cite{Witten:1992xu}
or \cite{Benini:2015noa,Benini:2016hjo,Closset:2015rna}.

Drawing from the cohomological functional (\ref{eq:V=000026Vprime})
and the field redefinitions (\ref{eq:dictStaVec}), we define a fermionic
functional of the fields of the $A$-model  vector multiplet
\begin{equation}
V(\tau,t)=\int\text{d}\mu\,\text{Tr}\,\left(-\frac{\tilde{\lambda}+\lambda}{4}\left(\tilde{D}-2i\star F\right)+\frac{i}{2}\lambda_{z}D_{\bar{z}}\tilde{\sigma}+\frac{i}{2}\lambda_{\bar{z}}D_{z}\tilde{\sigma}+\frac{i\tau\left(\tilde{\lambda}-\lambda\right)}{8}\left[\sigma,\tilde{\sigma}\right]-\frac{t\left(\tilde{\lambda}+\lambda\right)}{2}\tilde{\sigma}\right) ,\label{eq:V(tau,t)}
\end{equation}
where $\tau$ and $t$ are parameters. Although (\ref{eq:V(tau,t)})
is very similar to (\ref{eq:V=000026Vprime}), it has an additional
$\tau$ dependent part that permits a bosonic scalar commutator term. 

Acting with $Q_{A}$ on $V(\tau,t)$, we obtain a general $Q_{A}$-exact
functional
\begin{equation}
\begin{array}{ccl}
S(\tau,t) & = & Q_{A}V(\tau,t)\\
 & = & \int\text{d}\mu\,\text{Tr}\,\Bigg(\frac{1}{2}\left(\star F+it\tilde{\sigma}\right)^{2}+\frac{1}{2}\left(D+t\tilde{\sigma}\right)^{2}+\frac{1}{2}D_{\mu}\tilde{\sigma}D^{\mu}\sigma\\
 &  & \qquad\qquad+2i\lambda_{z}D_{\bar{z}}\tilde{\lambda}-2i\lambda_{\bar{z}}D_{z}\lambda+i\lambda_{\bar{z}}\left[\tilde{\sigma},\lambda_{z}\right]-\frac{i}{2}\tilde{\lambda}[\sigma,\lambda]-\frac{i}{4}\tilde{\lambda}[\sigma,\tilde{\lambda}]-\frac{i}{4}\lambda[\sigma,\lambda]\\
 &  & \qquad\qquad+2t\tilde{\lambda}\lambda+\tau\left(\frac{1}{8}\left[\sigma,\tilde{\sigma}\right]^{2}-\frac{i}{2}\tilde{\lambda}[\sigma,\lambda]+\frac{i}{4}\tilde{\lambda}[\sigma,\tilde{\lambda}]+\frac{i}{4}\lambda[\sigma,\lambda]\right)\Bigg),
\end{array}\label{eq:QAV(tau,t)}
\end{equation}
where we have used $\tilde{D}=D+i\star F$ then completed the square
for $\star F$ and $D$. The $t$-dependent part of (\ref{eq:QAV(tau,t)})
is the superpotential action (\ref{eq:SWb}) with quadratic $\overline{\widetilde{W}}(\tilde{\sigma})=\text{Tr}\tilde{\sigma}^{2}$,
or equivalently, the $t\delta V^{\prime}$ part of the cohomological
action (\ref{eq:ScohLoc}). For $t>0$, there is a mass term for the
scalar fermions and mixing terms between the bosonic scalar, the gauge
field, and the auxiliary field. The $\tau$-dependent part of (\ref{eq:QAV(tau,t)})
is the $\frac{1}{8}\left[\sigma,\tilde{\sigma}\right]^{2}$ term together
with its superpartner $-\frac{i}{2}\tilde{\lambda}[\sigma,\lambda]+\frac{i}{4}\tilde{\lambda}[\sigma,\tilde{\lambda}]+\frac{i}{4}\lambda[\sigma,\lambda]$,
which appears in the $A$-model  Yang-Mills action (\ref{eq:Ssym}),
but not in the cohomological action (\ref{eq:ScohLoc}). 

The localizing term (\ref{eq:Ssym}) used in \cite{Benini:2015noa,Benini:2016hjo,Closset:2015rna}
is obtained by setting $\tau=1$ and $t=0$ in (\ref{eq:QAV(tau,t)}),
which yields

\begin{equation}
\begin{array}{ccl}
S(\tau=1,t=0) & = & S_{\text{SYM}}\end{array}.\label{eq:QAV(1,0)}
\end{equation}
This can be expressed either as a variation with respect to $Q_{A}$,
or with respect to $Q$ and $\widetilde{Q}$, that is 
\begin{equation}
S_{\text{SYM}}=Q_{A}V(\tau=1,t=0)=\widetilde{Q}Q\,\int\text{d}\mu\,\text{Tr}\,\left(\frac{i}{2}\tilde{\sigma}\star F+\frac{1}{2}\tilde{\lambda}\lambda\right) . \label{eq:eq:QAV(1,0)alt}
\end{equation}
Setting $t=0$ eliminates the bosonic mixing terms, the fermionic
mass term, and leads to a locus with fermionic zero modes.

The $A$-model  analog of the cohomological localizing term
(\ref{eq:ScohLoc}) is obtained by setting $\tau=0$ and retaining
$t$ in (\ref{eq:QAV(tau,t)}), which yields 
\begin{equation}
\begin{array}{ccl}
S(t) & = & S(\tau=0,t>0)\\
 & = & \int\text{d}\mu\,\text{Tr}\,\Bigg(\frac{1}{2}\left(\star F+it\tilde{\sigma}\right)^{2}+\frac{1}{2}\left(D+t\tilde{\sigma}\right)^{2}+\frac{1}{2}D_{\mu}\tilde{\sigma}D^{\mu}\sigma\\
 &  & \qquad\qquad+2i\lambda_{z}D_{\bar{z}}\tilde{\lambda}-2i\lambda_{\bar{z}}D_{z}\lambda+i\lambda_{\bar{z}}\left[\tilde{\sigma},\lambda_{z}\right]-\frac{i}{2}\tilde{\lambda}[\sigma,\lambda]-\frac{i}{4}\tilde{\lambda}[\sigma,\tilde{\lambda}]-\frac{i}{4}\lambda[\sigma,\lambda]+2t\tilde{\lambda}\lambda\Bigg) .
\end{array}\label{eq:QAV(0,t)}
\end{equation}
For $t>0$ the bosonic mixing term between $\star F$ and $\tilde{\sigma}$
is retained. As we will see, this term conspires with the bosonic
scalar kinetic term to produce the Yang-Mills equations on the localization
locus. Furthermore, the presence of a fermionic mass term lifts the
fermionic zero modes, and the resulting localization locus is a bosonic
space rather than a superspace. Essentially, (\ref{eq:QAV(0,t)})
is the standard $A$-model  Yang-Mills action (\ref{eq:Ssym})
with the bosonic scalar commutator term and its superpartner removed,
plus a $Q_{A}$-exact twisted chiral superpotential (\ref{eq:SWb})
with quadratic $t$-dependent $\overline{\widetilde{W}}(\tilde{\sigma})=t\text{Tr}\tilde{\sigma}^{2}$.
Alternatively, (\ref{eq:QAV(0,t)}) with $t=0$ is a \textquotedblleft standard\textquotedblright{}
action for the two-dimensional analog of Donaldson theory \cite{Witten:1992xu}. 

\subsection{Localization locus}

In this section, we determine the localization locus -- the space
of field configurations for which the localizing term vanishes along
a particular integration contour. We place the coupling $1/h^{2}$
in front of the localizing term (\ref{eq:QAV(0,t)}) and consider localization in
the $h\to0$ limit. The bosonic part of the localizing term is 
\begin{equation}
\frac{1}{h^{2}}\int\text{d}\mu\;\text{Tr}\left(\frac{1}{2}\left(\star F+it\tilde{\sigma}\right)^{2}+\frac{1}{2}\left(D+t\tilde{\sigma}\right)^{2}+\frac{1}{2}D_{\mu}\tilde{\sigma}D^{\mu}\sigma+\cdots\right).\label{eq:locus1-1}
\end{equation}
The fields are valued in the adjoint representation of the Lie algebra
$\mathfrak{g}=\text{Lie}\,G$, which consists of hermitian matrices.
The positive definite metric on $\mathfrak{g}$ is $(a,b)=\text{Tr}\,ab$.
In the basis of hermitian generators, $T_{a}$, the fields are expressed
as $A_{\mu}=A_{\mu}^{a}T_{a}$, $\sigma=\sigma^{a}T_{a}$, $D_{\mu}\sigma=(D_{\mu}\sigma)^{a}T_{a}$,
and so on. 

The fields are generically complex in Euclidean signature,
in the sense that $\tilde{\sigma}$ is \emph{not} the hermitian conjugate
of $\sigma$. An integration contour is chosen for the path integral
by imposing reality conditions on the fields. Any contour for which
the path integral converges is acceptable. The integration contour
we choose closely resembles the one employed in section 3.2.2 of \cite{Witten:1992xu},
with the fields of the standard cohomological multiplet rewritten
in terms of the $A$-model vector multiplet using (\ref{eq:dictStaVec}). 

To begin, the auxiliary field $D$ is integrated out by setting it
to its on-shell value $D=-t\tilde{\sigma}$ in (\ref{eq:locus1-1}),
eliminating the $\frac{1}{2}(D+t\tilde{\sigma})^{2}$ term. We then
solve the matrix equation
\begin{equation}
0=\text{Tr}\left(\star F+it\tilde{\sigma}\right)^{2}+\text{Tr}D_{\mu}\tilde{\sigma}D^{\mu}\sigma\label{eq:locus2-1}
\end{equation}
for the fields $A_{\mu}$,\;$\sigma$, and $\tilde{\sigma}$. Notice
that $A_{\mu}$ appears in both $\star F$ and $D_{\mu}$. The gauge
field $A_{\mu}$ is taken to be real, which fixes $\star F$ to be
real. Typically, the bosonic scalars $\sigma$ and $\tilde{\sigma}$
are taken to be hermitian conjugates of one another, i.e. $\tilde{\sigma}=\sigma^{\dagger}$.
However, this choice is incompatible with the real $\star F$ in (\ref{eq:locus2-1}).
Instead, we take $\sigma$ to be real, $\tilde{\sigma}$ to be purely
imaginary, i.e., $\tilde{\sigma}=i\tilde{\sigma}_{E}$, and $t$ to
be real. In the following sections, we will use $\tilde{\sigma}=\iota\tilde{\sigma}_{E}$,
and set $\iota=i$ at the end. For real $\sigma$ and imaginary $\tilde{\sigma}=i\tilde{\sigma}_{E}$, equation (\ref{eq:locus2-1}) reads 
\begin{equation}
0=\text{Tr}\left(\star F-t\tilde{\sigma}_{E}\right)^{2}+i\text{Tr}D_{\mu}\tilde{\sigma}_{E}D^{\mu}\sigma .\label{eq:locus3-1}
\end{equation}
The first term vanishes because it is a trace over squared hermitian
matrices. The second term vanishes by requiring the integral over
the bosonic scalars to have stationary phase. Consequently, (\ref{eq:locus3-1})
reduces to 
\begin{equation}
\star F-t\tilde{\sigma}_{E}=0,\qquad D_{\mu}\tilde{\sigma}_{E}=0,\qquad D_{\mu}\sigma=0.\label{eq:locus4-1}
\end{equation}
Combining the first two equations, we obtain the Yang-Mills equation
on $S^{2}$,
\begin{equation}
0=\frac{1}{t}D_{\mu}\star F,\label{eq:locus5-1}
\end{equation}
for $t>0$. Due to the singularity at $t=0$, the solutions are said
to “flow in from infinity” for $t>0$ \cite{Witten:1992xu}. The solutions
of the Yang-Mills equations on a closed Riemannian manifold of genus
$g$ were classified in \cite{Atiyah:1982fa}. The analysis here focuses
on the genus zero case of $S^{2}$.

The solutions of (\ref{eq:locus5-1}) are described by two types of
connections $A=A_{\mu}\,\text{d}x^{\mu}$: flat connections $A$ with
zero curvature $F=0$ and Yang-Mills connections $A$ with non-zero
curvature $F\neq0$. At genus zero, the flat connections are gauge-equivalent
to the trivial connection $A=0$. In contrast, the flux of a Yang-Mills
connection $A$ is GNO quantized,
\begin{equation}
\frac{1}{2\pi}\int_{S^{2}}F=\mathfrak{m}\in\Lambda_{\text{cochar}},\label{eq:locus6-1}
\end{equation}
where $\Lambda_{\text{cochar}}=\left\{ \gamma\in\mathfrak{h}\mid e^{2\pi i\gamma}=1_{G}\right\} $
is the cocharacter lattice in the Cartan subalgebra $\mathfrak{h}$
of $\mathfrak{g}$, and $1_{G}$ is the identity element in $G$.
In particular, the locus (\ref{eq:locus4-1}) includes the Yang-Mills
equation for $t>0$, which in turn ensures the GNO quantization of
the gauge flux. Given the quantization of the flux, the first two
equations of the locus reduce to 
\begin{equation}
\star F=\frac{\mathfrak{m}}{2R^{2}},\qquad\tilde{\sigma}_{E}=\frac{\mathfrak{m}}{2tR^{2}}.\label{eq:locus7-1}
\end{equation}
Consequently, the gauge field $A_{\mu}$ and the bosonic scalar $\tilde{\sigma}_{E}$
are parameterized by the discrete modulus $\mathfrak{m}\in\Lambda_{\text{cochar}}$
in $\mathfrak{h}$.

Next, we consider the third equation of (\ref{eq:locus4-1}), 
\begin{equation}
0=D_{\mu}\sigma=\partial_{\mu}\sigma-i[A_{\mu},\sigma],\label{eq:locus8-1}
\end{equation}
whose solutions depend on whether the connection $A$ is reducible
or irreducible\footnote{The connection $A$ is called reducible if $\text{stab}_{G}(A)\neq Z(G)$
and irreducible if $\text{stab}_{G}(A)=Z(G)$, where $G$ is the gauge
group, $\text{stab}_{G}(A)$ is the stabilizer group of $A$, and
$Z(G)$ is the center of $G$.} \cite{Deligne:1999qp}. If $A$ is irreducible, $D_{\mu}\sigma=0$
is solved by $\sigma=0$. If $A$ is reducible, $D_{\mu}\sigma=0$
is solved by non-zero $\sigma$ that satisfies $0=\partial_{\mu}\sigma=[A_{\mu},\sigma]$,
i.e., constant configurations of $\sigma$ that commute with $A_{\mu}$.
At genus zero, the flat connections are irreducible and gauge-equivalent to the trivial connection, while Yang-Mills connections are
reducible. Our derivation focuses on non-trivial solutions of (\ref{eq:locus5-1})
and (\ref{eq:locus8-1}) in which $\sigma$ is a non-zero constant
configuration and $A_{\mu}$ is a reducible Yang-Mills connection.

Acting on (\ref{eq:locus8-1}) with covariant
derivatives, we obtain 
\begin{equation}
0=[F_{\mu\nu},\sigma].\label{eq:locus9}
\end{equation}
Given that $\sigma$ commutes with $F_{\mu\nu}$, $\sigma$ can be
diagonalized (conjugated into the Cartan subalgebra $\mathfrak{h}$)
using an element of $G$. Consequently, $\sigma$ is parameterized
by the real continuous modulus $u=-2\sigma$ in 
$\mathfrak{h}$. In summary, the fields $A_{\mu}$, $\sigma$, and
$\tilde{\sigma}_{E}$ are parameterized by the moduli $u$ and $\mathfrak{m}$
according to:
\begin{equation}
\star F=\frac{\mathfrak{m}}{2R^{2}},\qquad\tilde{\sigma}_{E}=\frac{\mathfrak{m}}{2tR^{2}},\qquad\sigma=-\frac{u}{2}.\label{eq:locus10}
\end{equation}
Notice that these field configurations are also the bosonic fixed-points
of the fermionic symmetry, in the sense that they are solutions of
\begin{equation}
\begin{array}{lcccl}
0 & = & Q_{A}\lambda_{z} & = & -2iD_{z}\sigma,\\
0 & = & Q_{A}\lambda & = & -D-i\star F+\frac{i}{2}\left[\sigma,\tilde{\sigma}\right],\\
0 & = & Q_{A}\tilde{\lambda} & = & -D-i\star F-\frac{i}{2}\left[\sigma,\tilde{\sigma}\right],\\
0 & = & Q_{A}\lambda_{\bar{z}} & = & -2iD_{\bar{z}}\sigma.
\end{array}\label{eq:locus11}
\end{equation}

\subsection{Gauge-fixing}

The localizing term is gauge-fixed according to the standard BRST
procedure. We construct a gauge-fixing action that will be added to
the localizing term, resulting in the gauge-fixed localizing term.
The gauge-fixing action is closed under a linear combination of the
BRST symmetry generator and the $A$-model  scalar supercharge.
The gauge-fixed localizing term will be the quantity that we expand
around the localization locus in the next section. 

We introduce the BRST operator $Q_{\mathbf{B}}$, the anti-commuting
scalar ghosts $c,\tilde{c}$, and a commuting scalar $B$. The action
of $Q_{\mathbf{B}}$ on the fields is 
\begin{equation}
\begin{array}{lcccc}
Q_{\mathbf{B}}\,A_{\mu}=D_{\mu}c, &  & Q_{\mathbf{B}}\,\tilde{c}=B, &  & Q_{\mathbf{B}}\,\Phi=i[c,\Phi],\\
\\
Q_{\mathbf{B}}\,c=i\left\{ c,c\right\} , &  & Q_{\mathbf{B}}\,B=0,
\end{array}\label{eq:QBtransf}
\end{equation}
where $\Phi$ denotes $\sigma,\tilde{\sigma},\tilde{D},\lambda_{\mu},\lambda,\tilde{\lambda}$. 

Next, we define the linear combination of the $A$-model  and
BRST operators 
\begin{equation}
\hat{Q}=Q_{A}+Q_{\mathbf{B}}.
\end{equation}
To determine the action of $\hat{Q}$ on the fields, we must specify
the action of $Q_{A}$ on $c,\tilde{c},B$. All other transformations
are described in (\ref{eq:QAtransf}) and (\ref{eq:QBtransf}). At
this point, one could introduce ghosts for ghosts and go on to describe
the combined BRST-supersymmetry complex in full detail, along the
lines of section 4.1 in \cite{Pestun:2007rz}. However, we restrict
ourselves to a simpler resolution. As noted in \cite{Ohta:2019odi},
a reasonable choice is to assume that $Q_{A}$ acts on $c,\tilde{c},B$
as 
\begin{eqnarray}
Q_{A}c & = & -\sigma,\\
Q_{A}\tilde{c} & = & 0,\\
Q_{A}B & = & 0.
\end{eqnarray}
In other words, we assume that the bosonic scalar $\sigma$ is the
superpartner of the ghost $c$. Consequently, the action of $\hat{Q}$
on the fields is nilpotent 
\begin{eqnarray}
0 & = & \hat{Q}^{2}A_{\mu}=D_{\mu}\left(\sigma+Q_{A}c\right),\\
0 & = & \hat{Q}^{2}\sigma=i\left[Q_{A}c,\sigma\right],\\
0 & = & \hat{Q}^{2}\Phi=i\left[\Phi,\sigma+Q_{A}c\right],
\end{eqnarray}
where $\Phi$ denotes $\tilde{\sigma},\tilde{D},\lambda_{\mu},\lambda,\tilde{\lambda}$.

The $\hat{Q}$-exact gauge-fixing Lagrangian is
\[
L_{\text{gf}}=p_{1}\,\hat{Q}\,\text{Tr}\,\left(\tilde{c}\,\left(p_{2}\,G_{\text{gf}}+p_{3}\,B\right)\right)
\]
where $G_{\text{gf}}$ is the gauge-fixing function and $p_{1},p_{2},p_{3}$
are parameters. Acting with $\hat{Q}$, then completing the square
for $B$, we obtain 
\[
L_{\text{gf}}=\text{Tr}\left(-\frac{p_{2}^{2}p_{1}}{4p_{3}}G_{\text{gf}}^{2}+p_{1}p_{3}\left(B+\frac{p_{2}}{2p_{3}}G_{\text{gf}}\right)^{2}-p_{2}p_{1}\tilde{c}\hat{Q}G_{\text{gf}}\right)
\]
The gauge fixing function is taken to be 
\begin{equation}
G_{\text{gf}}=D_{\mu}^{0}A^{\mu}=2\left(D_{z}^{0}A_{\bar{z}}+D_{\bar{z}}^{0}A_{z}\right),
\end{equation}
where $D_{\mu}^{0}=\partial_{\mu}-iA_{\mu}^{0}$ and $A_{\mu}^{0}$
is valued on the locus (\ref{eq:locus4-1}). Integrating out $B$, and
setting the parameters to $p_{1}=1,\,p_{2}=i,\,p_{3}=\frac{\xi}{2}$
, the gauge-fixing Lagrangian is
\begin{eqnarray}
L_{\text{gf}} & = & \text{Tr}\bigg(\frac{1}{2\xi}\left(D_{\mu}^{0}A^{\mu}\right)^{2}-i\tilde{c}D_{\mu}^{0}D_{0}^{\mu}c+2\tilde{c}D_{z}^{0}\lambda_{\bar{z}}+2\tilde{c}D_{\bar{z}}^{0}\lambda_{z}\bigg)\label{eq:Lgf}
\end{eqnarray}
Adding (\ref{eq:Lgf}) to (\ref{eq:QAV(0,t)}), we obtain the gauge-fixed
localizing term 
\begin{equation}
\begin{array}{ccl}
S_{\text{loc}} & = & \frac{1}{h^{2}}(S(t)+S_{\text{gf}})\\
\\
 & = & \frac{1}{h^{2}}\,\int\text{d}\mu\,\text{Tr}\,\bigg[\frac{1}{2}\left(\star F+it\iota\tilde{\sigma}_{E}\right)^{2}+\frac{\iota}{2}D_{\mu}\tilde{\sigma}_{E}D^{\mu}\sigma\\
\\
 &  & \qquad\qquad\quad+2i\tilde{\lambda}D_{\bar{z}}\lambda_{z}-2i\lambda_{\bar{z}}D_{z}\lambda+i\iota\lambda_{\bar{z}}\left[\tilde{\sigma}_{E},\lambda_{z}\right]-\frac{i}{4}\tilde{\lambda}[\sigma,\tilde{\lambda}]-\frac{i}{4}\lambda[\sigma,\lambda]\\
\\
 &  & \qquad\qquad\quad-\frac{i}{2}\tilde{\lambda}[\sigma,\lambda]+2t\tilde{\lambda}\lambda+\frac{1}{2\xi}\left(D_{\mu}^{0}A^{\mu}\right)^{2}+2\tilde{c}D_{z}^{0}\lambda_{\bar{z}}+2\tilde{c}D_{\bar{z}}^{0}\lambda_{z}-i\tilde{c}D_{\mu}^{0}D_{0}^{\mu}c\bigg]
\end{array}\label{eq:Sloc0}
\end{equation}

\subsection{Locus expansion}

Here we describe the expansion of the gauge-fixed localizing term
around the localization locus, then construct the fluctuation operators.
The fields are separated into zero modes and fluctuating modes, where
the zero modes are valued on the locus (\ref{eq:locus4-1}). Then, we
perform a functional Taylor series expansion of the localizing term
around the locus configurations. Following this, we take the localizing
limit $h\to0$, such that the expanded localizing term only retains
terms that are quadratic in fluctuations. 

Explicitly, the fields are decomposed as 
\begin{equation}
\varphi=\varphi_{0}+h\,\varphi^{\prime}=\sum_{k=1}^{\text{rank}G}\varphi_{0}^{k}\,\mathtt{H}_{k}+h\,\left(\sum_{k=1}^{\text{rank}G}\varphi^{\prime k}\,\mathtt{H}_{k}+\sum_{\alpha\in\Delta}\varphi^{\prime\alpha}\,\mathtt{E}_{\alpha}\right),\label{eq:CWdecomp}
\end{equation}
where $\varphi$ is a field of the vector multiplet, $\varphi_{0}$
is a zero-mode, $\varphi^{\prime}$ is a fluctuating mode, $\alpha$
are roots, $\Delta$ is the set of roots, $k$ is a Cartan index, and
$\left\{ \mathtt{H}_{k},\mathtt{E}_{\alpha}\right\} $ is the Cartan-Weyl
basis. The bosonic zero modes $A_{\mu,0},\sigma_{0},\tilde{\sigma}_{0}$
are given by (\ref{eq:locus10}). The fermionic and ghost zero-modes
$\lambda_{\mu,0},\lambda_{0},\tilde{\lambda}_{0},c_{0},\tilde{c}_{0}$
are set to zero. 

Decomposing the fields in the gauge-fixed localizing term (\ref{eq:Sloc0})
according to (\ref{eq:CWdecomp}), then taking the localizing limit
$h^{2}\to0$, we find the gauge-fixed localizing term to quadratic
order in fluctuations 
\begin{equation}
\begin{array}{ccl}
\tilde{S}_{\text{loc}} & = & \int\text{d}\mu\,\text{Tr}\bigg(\frac{1}{2}\left(\star F^{\prime}+it\iota\tilde{\sigma}_{E}^{\prime}\right)^{2}+\frac{\iota}{2}\left(D_{\mu}^{0}\tilde{\sigma}_{E}^{\prime}-i\left[A_{\mu}^{\prime},\tilde{\sigma}_{E,0}\right]\right)\left(D_{0}^{\mu}\sigma^{\prime}-i\left[A^{\prime\mu},\sigma_{0}\right]\right)\\
\\
 &  & \qquad\qquad+2i\tilde{\lambda}^{\prime}D_{\bar{z}}^{0}\lambda_{z}^{\prime}-2i\lambda_{\bar{z}}^{\prime}D_{z}^{0}\lambda^{\prime}+i\iota\lambda_{\bar{z}}^{\prime}\left[\tilde{\sigma}_{E,0},\lambda_{z}^{\prime}\right]-\frac{i}{4}\tilde{\lambda}^{\prime}[\sigma_{0},\tilde{\lambda}^{\prime}]-\frac{i}{4}\lambda^{\prime}[\sigma_{0},\lambda^{\prime}]\\
\\
 &  & \qquad\qquad-\frac{i}{2}\tilde{\lambda}^{\prime}[\sigma_{0},\lambda^{\prime}]+2t\tilde{\lambda}^{\prime}\lambda^{\prime}+\frac{1}{2\xi}\left(D_{\mu}^{0}A^{\prime\mu}\right)^{2}+2\tilde{c}^{\prime}D_{z}^{0}\lambda_{\bar{z}}^{\prime}+2\tilde{c}^{\prime}D_{\bar{z}}^{0}\lambda_{z}^{\prime}-i\tilde{c}^{\prime}D_{\mu}^{0}D_{0}^{\mu}c^{\prime}\,\bigg)+\mathcal{O}\left(h\right)
\end{array}\label{eq:LlocTilde}
\end{equation}
where $\star F^{\prime}=-2i\left(D_{z}^{0}A_{\bar{z}}^{\prime}-D_{\bar{z}}^{0}A_{z}^{\prime}\right)$
and $D_{\mu}^{0}=\partial_{\mu}-iA_{\mu}^{0}$. We drop the prime
on fluctuations to keep notation clean. 

Introducing the basis 
\begin{equation}
\begin{array}{lcl}
\bar{\Phi}^{k}=(A_{\bar{z}}^{k},A_{z}^{k},\sigma^{k},\tilde{\sigma}_{E}^{k}), &  & \Phi^{k}=(A_{z}^{k},A_{\bar{z}}^{k},\sigma^{k},\tilde{\sigma}_{E}^{k})^{\top},\\
\bar{\Psi}^{k}=(\lambda_{\bar{z}}^{k},\lambda_{z}^{k},\tilde{\lambda}^{k},\lambda^{k},\tilde{c}^{k},c^{k}), &  & \Psi^{k}=(\lambda_{z}^{k},\lambda_{\bar{z}}^{k},\lambda^{k},\tilde{\lambda}^{k},c^{k},\tilde{c}^{k})^{\top},\\
\bar{\Phi}^{-\alpha}=(A_{\bar{z}}^{-\alpha},A_{z}^{-\alpha},\sigma^{-\alpha},\tilde{\sigma}_{E}^{-\alpha}), &  & \Phi^{\alpha}=(A_{z}^{\alpha},A_{\bar{z}}^{\alpha},\sigma^{\alpha},\tilde{\sigma}_{E}^{\alpha})^{\top},\\
\bar{\Psi}^{-\alpha}=(\lambda_{\bar{z}}^{-\alpha},\lambda_{z}^{-\alpha},\tilde{\lambda}^{-\alpha},\lambda^{-\alpha},\tilde{c}^{-\alpha},c^{-\alpha}), &  & \Psi^{\alpha}=(\lambda_{z}^{\alpha},\lambda_{\bar{z}}^{\alpha},\lambda^{\alpha},\tilde{\lambda}^{\alpha},c^{\alpha},\tilde{c}^{\alpha})^{\top},
\end{array}\label{eq:basis}
\end{equation}
the quadratic action (\ref{eq:LlocTilde}) can be written in terms
of fluctuation operators
\begin{equation}
\tilde{S}_{\text{loc}}  =  \int\text{d}\mu\,\text{Tr}\bigg(\sum_{k=1}^{\text{rank}G}\left(\bar{\Phi}^{k}\Delta_{\mathrm{b}}\Phi^{k}+\bar{\Psi}^{k}\Delta_{\mathrm{f}}\Psi^{k}\right)+\sum_{\alpha\in\Delta}\left(\bar{\Phi}^{-\alpha}\Delta_{\mathrm{B}}\Phi^{\alpha}+\bar{\Psi}^{-\alpha}\Delta_{\mathrm{F}}\Psi^{\alpha}\right)\bigg)+\mathcal{O}\left(h\right).\label{eq:LlocTildeMatrix}
\end{equation}
The operators in the Cartan part of (\ref{eq:LlocTildeMatrix}) are
\begin{equation}
\begin{array}{ccl}
\Delta_{\mathrm{b}} & = & \left(\begin{array}{cccc}
-\frac{2(\xi+1)D_{z}^{0}D_{\bar{z}}^{0}}{\xi} & \frac{2(\xi-1)D_{z}^{0}D_{z}^{0}}{\xi} & 0 & -itD_{z}^{0}\\
\frac{2(\xi-1)D_{\bar{z}}^{0}D_{\bar{z}}^{0}}{\xi} & -\frac{2(\xi+1)D_{\bar{z}}^{0}D_{z}^{0}}{\xi} & 0 & itD_{\bar{z}}^{0}\\
0 & 0 & 0 & -\frac{i}{4}D_{\mu}^{0}D_{0}^{\mu}\\
-itD_{\bar{z}}^{0} & itD_{z}^{0} & -\frac{i}{4}D_{\mu}^{0}D_{0}^{\mu} & \frac{t^{2}}{2}
\end{array}\right)\\
\\
\Delta_{\mathrm{f}} & = & \left(\begin{array}{cccccc}
0 & 0 & -iD_{z}^{0} & 0 & 0 & D_{z}^{0}\\
0 & 0 & 0 & iD_{\bar{z}}^{0} & 0 & D_{\bar{z}}^{0}\\
iD_{\bar{z}}^{0} & 0 & t & 0 & 0 & 0\\
0 & -iD_{z}^{0} & 0 & -t & 0 & 0\\
D_{\bar{z}}^{0} & D_{z}^{0} & 0 & 0 & iD_{\mu}^{0}D_{0}^{\mu} & 0\\
0 & 0 & 0 & 0 & 0 & -iD_{\mu}^{0}D_{0}^{\mu}
\end{array}\right)
\end{array}
\end{equation}
The operators in the root space part of (\ref{eq:LlocTildeMatrix})
are 
\begin{equation}
\begin{array}{ccl}
\Delta_{\mathrm{B}} & = & \left(\begin{array}{cccc}
-\frac{2(\xi+1)D_{z}^{0}D_{\bar{z}}^{0}}{\xi}+\iota\alpha\left(\sigma\right)\alpha\left(\tilde{\sigma}_{E}\right) & \frac{2(\xi-1)D_{z}^{0}D_{z}^{0}}{\xi} & -\frac{i\iota\alpha\left(\tilde{\sigma}_{E}\right)D_{z}^{0}}{2} & -\frac{\iota\left(2t+i\alpha\left(\sigma\right)\right)D_{z}^{0}}{2}\\
\frac{2(\xi-1)D_{\bar{z}}^{0}D_{\bar{z}}^{0}}{\xi} & -\frac{2(\xi+1)D_{\bar{z}}^{0}D_{z}^{0}}{\xi}+\iota\alpha\left(\sigma\right)\alpha\left(\tilde{\sigma}_{E}\right) & -\frac{i\iota\alpha\left(\tilde{\sigma}_{E}\right)D_{\bar{z}}^{0}}{2} & \frac{\iota\left(2t-i\alpha\left(\sigma\right)\right)D_{\bar{z}}^{0}}{2}\\
-\frac{i\iota\alpha\left(\tilde{\sigma}_{E}\right)D_{\bar{z}}^{0}}{2} & -\frac{i\iota\alpha\left(\tilde{\sigma}_{E}\right)D_{z}^{0}}{2} & 0 & -\frac{\iota D_{\mu}^{0}D_{0}^{\mu}}{4}\\
-\frac{\iota\left(2t+i\alpha\left(\sigma\right)\right)D_{\bar{z}}^{0}}{2} & \frac{\iota\left(2t-i\alpha\left(\sigma\right)\right)D_{z}^{0}}{2} & -\frac{\iota D_{\mu}^{0}D_{0}^{\mu}}{4} & -\frac{t^{2}\iota}{2}
\end{array}\right),\\
\\
\Delta_{\mathrm{F}} & = & \left(\begin{array}{cccccc}
-\frac{\alpha\left(\tilde{\sigma}_{E}\right)}{2} & 0 & -iD_{z}^{0} & 0 & 0 & D_{z}^{0}\\
0 & -\frac{\alpha\left(\tilde{\sigma}_{E}\right)}{2} & 0 & iD_{\bar{z}}^{0} & 0 & D_{\bar{z}}^{0}\\
iD_{\bar{z}}^{0} & 0 & t-\frac{i\alpha\left(\sigma\right)}{4} & -\frac{i\alpha\left(\sigma\right)}{4} & 0 & 0\\
0 & -iD_{z}^{0} & -\frac{i\alpha\left(\sigma\right)}{4} & -t-\frac{i\alpha\left(\sigma\right)}{4} & 0 & 0\\
D_{\bar{z}}^{0} & D_{z}^{0} & 0 & 0 & \frac{i}{2}D_{\mu}^{0}D_{0}^{\mu} & 0\\
0 & 0 & 0 & 0 & 0 & -\frac{i}{2}D_{\mu}^{0}D_{0}^{\mu}
\end{array}\right),
\end{array}
\end{equation}
where the roots are $\alpha(\sigma)=\alpha(\mathtt{H}^{k})\sigma_{0}^{k}$
and $\alpha(\tilde{\sigma}_{E})=\alpha(\mathtt{H}^{k})\tilde{\sigma}_{E,0}^{k}$.
In what follows, we use (\ref{eq:locus10}) to express the roots
in terms of moduli, $\alpha(\sigma)=-\frac{1}{2}\alpha(u)$ and $\alpha(\tilde{\sigma}_{E})=\frac{i}{2\iota R^{2}t}\alpha(\mathfrak{m})$.
Then, we set $\iota=i$ and choose the gauge $\xi=1$. Furthermore,
we will scale $c^{\alpha}$ and $\tilde{c}^{\alpha}$ by a factor
of $\sqrt{2}$.

\subsection{One-loop determinant}

To evaluate the one-loop determinant mode-by-mode, we express the
quadratic localizing term in a basis of monopole spherical harmonics,
evaluate the contribution of each mode, collect the contributions,
then present the full one-loop determinant. The gauge-fixing turns
out to be singular in the $\alpha(\mathfrak{m})=0$ sector, and we
gauge-fix once more following the approach of \cite{Blau:1993hj}. 

We use the monopole spherical harmonics of \cite{Benini:2012ui,Benini:2015noa},
which are 
\begin{equation}
Y_{j,j_{3}}^{s}\text{ for }j\geq\left|s\right|,\qquad Y_{j,j_{3}}^{s+1}\text{ for }j\geq\left|s+1\right|,\qquad Y_{j,j_{3}}^{s-1}\text{ for }j\geq\left|s-1\right|,\label{eq:harmonics}
\end{equation}
where $s=-\frac{1}{2}c_{1}$ is the effective spin and $c_{1}$ is
the first Chern number of the line bundle. The harmonics obey 
\begin{equation}
\begin{array}{ccccl}
D_{z}^{0}Y_{j,j_{3}}^{s}=\frac{s_{+}}{2R}Y_{j,j_{3}}^{s+1}, &  & D_{z}^{0}D_{\bar{z}}^{0}Y_{j,j_{3}}^{s+1}=-\frac{s_{+}^{2}}{4R^{2}}Y_{j,j_{3}}^{s+1}, &  & D_{z}^{0}D_{z}^{0}Y_{j,j_{3}}^{s-1}=\frac{s_{-}s_{+}}{4R^{2}}Y_{j,j_{3}}^{s-1},\\
\\
D_{\bar{z}}^{0}Y_{j,j_{3}}^{s}=-\frac{s_{-}}{2R}Y_{j,j_{3}}^{s-1}, &  & D_{\bar{z}}^{0}D_{z}^{0}Y_{j,j_{3}}^{s-1}=-\frac{s_{-}^{2}}{4R^{2}}Y_{j,j_{3}}^{s-1}, &  & D_{\bar{z}}^{0}D_{\bar{z}}^{0}Y_{j,j_{3}}^{s+1}=\frac{s_{-}s_{+}}{4R^{2}}Y_{j,j_{3}}^{s+1},\\
\\
D_{\bar{z}}^{0}Y_{j,j_{3}}^{s+1}=-\frac{s_{+}}{2R}Y_{j,j_{3}}^{s}, &  & D_{z}^{0}D_{\bar{z}}^{0}Y_{j,j_{3}}^{s}=-\frac{s_{-}^{2}}{4R^{2}}Y_{j,j_{3}}^{s}, &  & D_{z}^{0}Y_{j,j_{3}}^{s+1}=0,\\
\\
D_{z}^{0}Y_{j,j_{3}}^{s-1}=\frac{s_{-}}{2R}Y_{j,j_{3}}^{s}, &  & D_{\bar{z}}^{0}D_{z}^{0}Y_{j,j_{3}}^{s}=-\frac{s_{+}^{2}}{4R^{2}}Y_{j,j_{3}}^{s}, &  & D_{\bar{z}}^{0}Y_{j,j_{3}}^{s-1}=0,
\end{array}\label{eq:eigenvalues}
\end{equation}
where
\begin{equation}
s_{\pm}=\sqrt{j(j+1)-s(s\pm1)}.
\end{equation}
To express the quadratic localizing term (\ref{eq:LlocTildeMatrix})
in the basis of harmonics (\ref{eq:harmonics}), the fields $\Phi^{\alpha}=(A_{z}^{\alpha},A_{\bar{z}}^{\alpha},\sigma^{\alpha},\tilde{\sigma}_{E}^{\alpha})^{\top}$
are expanded in terms of the harmonics $(Y_{j,j_{3}}^{s+1},Y_{j,j_{3}}^{s-1},Y_{j,j_{3}}^{s},Y_{j,j_{3}}^{s})^{\top}$,
and the fields $\Psi^{\alpha}=(\lambda_{z}^{\alpha},\lambda_{\bar{z}}^{\alpha},\lambda^{\alpha},\tilde{\lambda}^{\alpha},c^{\alpha},\tilde{c}^{\alpha})^{\top}$
are expanded in terms of the harmonics $(Y_{j,j_{3}}^{s+1},Y_{j,j_{3}}^{s-1},Y_{j,j_{3}}^{s},Y_{j,j_{3}}^{s},Y_{j,j_{3}}^{s},Y_{j,j_{3}}^{s})^{\top}$.
The bosonic and fermionic operators acting on harmonics are
\begin{equation}
\Delta_{\text{B}}\Phi^{\alpha}=\left(\begin{array}{cccc}
-4D_{z}^{0}D_{\bar{z}}^{0}-\frac{i\alpha(\mathfrak{m})\alpha(u)}{4R^{2}t} & 0 & \frac{\alpha(\mathfrak{m})D_{z}^{0}}{4R^{2}t} & \left(-\frac{\alpha(u)}{4}-it\right)D_{z}^{0}\\
0 & -4D_{\bar{z}}^{0}D_{z}^{0}-\frac{i\alpha(\mathfrak{m})\alpha(u)}{4R^{2}t} & \frac{\alpha(\mathfrak{m})D_{\bar{z}}^{0}}{4R^{2}t} & \left(-\frac{\alpha(u)}{4}+it\right)D_{\bar{z}}^{0}\\
\frac{\alpha(\mathfrak{m})}{4R^{2}t}D_{\bar{z}}^{0} & \frac{\alpha(\mathfrak{m})}{4R^{2}t}D_{z}^{0} & 0 & -\frac{i}{4}D_{\mu}^{0}D_{0}^{\mu}\\
\left(-\frac{\alpha(u)}{4}-it\right)D_{\bar{z}}^{0} & \left(-\frac{\alpha(u)}{4}+it\right)D_{z}^{0} & -\frac{i}{5}D_{\mu}^{0}D_{0}^{\mu} & \frac{t^{2}}{2}
\end{array}\right)\left(\begin{array}{c}
Y_{j,j_{3}}^{s+1}\\
Y_{j,j_{3}}^{s-1}\\
Y_{j,j_{3}}^{s}\\
Y_{j,j_{3}}^{s}
\end{array}\right),\label{eq:Obos}
\end{equation}
\begin{equation}
\Delta_{\text{F}}\Psi^{\alpha}=\left(\begin{array}{cccccc}
-\frac{\alpha(\mathfrak{m})}{4R^{2}t} & 0 & -iD_{z}^{0} & 0 & 0 & \sqrt{2}D_{z}^{0}\\
0 & -\frac{\alpha(\mathfrak{m})}{4R^{2}t} & 0 & iD_{\bar{z}}^{0} & 0 & \sqrt{2}D_{\bar{z}}^{0}\\
iD_{\bar{z}}^{0} & 0 & t+\frac{i\alpha(u)}{8} & \frac{i\alpha(u)}{8} & 0 & 0\\
0 & -iD_{z}^{0} & \frac{i\alpha(u)}{8} & -t+\frac{i\alpha(u)}{8} & 0 & 0\\
\sqrt{2}D_{\bar{z}}^{0} & \sqrt{2}D_{z}^{0} & 0 & 0 & iD_{\mu}^{0}D_{0}^{\mu} & 0\\
0 & 0 & 0 & 0 & 0 & -iD_{\mu}^{0}D_{0}^{\mu}
\end{array}\right)\left(\begin{array}{c}
Y_{j,j_{3}}^{s+1}\\
Y_{j,j_{3}}^{s-1}\\
Y_{j,j_{3}}^{s}\\
Y_{j,j_{3}}^{s}\\
Y_{j,j_{3}}^{s}\\
Y_{j,j_{3}}^{s}
\end{array}\right).\label{eq:Ofer}
\end{equation}
To evaluate the contribution of each mode, we first specify $j$,
determine which harmonics exist using (\ref{eq:harmonics}), and remove
rows and columns from (\ref{eq:Obos}) and (\ref{eq:Ofer}) if necessary.
Following this, we replace $D_{z}^{0}$ and $D_{\bar{z}}^{0}$ with
eigenvalues $s_{\pm}$ using (\ref{eq:eigenvalues}), then evaluate
the determinants and their ratio. 

For $j\geq\frac{|\alpha(\mathfrak{m})|}{2}+1,\;\alpha(\mathfrak{m})\in\mathbb{Z}$,
the harmonics $Y_{j,j_{3}}^{s-1},\;Y_{j,j_{3}}^{s},\;Y_{j,j_{3}}^{s+1}$
exist, and the eigenvalues have multiplicity $2j+1$. The ratio of
determinants is 
\begin{equation}
\frac{\sqrt{\det\Delta_{\mathrm{F}}}}{\sqrt{\det\Delta_{\mathrm{B}}}}  =  \prod_{\alpha\in\Delta}\prod_{j\geq\frac{|\alpha(\mathfrak{m})|}{2}+1}\prod_{j_{3}=-j}^{j}\frac{\left(\frac{\left(j^{2}+j-\frac{\alpha(\mathfrak{m})^{2}}{4}\right)^{2}\left(\left(j^{2}+j-\frac{\alpha(\mathfrak{m})^{2}}{4}\right)\left(j^{2}+j-\frac{\alpha(\mathfrak{m})}{2}\left(\frac{\alpha(\mathfrak{m})}{2}+\frac{i\alpha(u)}{t}\right)\right)-\frac{\alpha(\mathfrak{m})^{2}}{4}\right)}{16R^{8}}\right)^{1/2}}{\left(\frac{\left(j^{2}+j-\frac{\alpha(\mathfrak{m})^{2}}{4}\right)^{2}\left(\left(j^{2}+j-\frac{\alpha(\mathfrak{m})^{2}}{4}\right)\left(j^{2}+j-\frac{\alpha(\mathfrak{m})}{2}\left(\frac{\alpha(\mathfrak{m})}{2}+\frac{i\alpha(u)}{t}\right)\right)-\frac{\alpha(\mathfrak{m})^{2}}{4}\right)}{16R^{8}}\right)^{1/2}}.
\end{equation}
Switching to positive roots then evaluating the product over $j_{3}$
results in
\begin{equation}
\frac{\sqrt{\det\Delta_{\mathrm{F}}}}{\sqrt{\det\Delta_{\mathrm{B}}}}=\prod_{\alpha\in\Delta_{+}}\prod_{j\geq\frac{|\alpha(\mathfrak{m})|}{2}+1}\left(1\right)^{2j+1}
\end{equation}
Using $j=\ell+\frac{|\alpha(\mathfrak{m})|}{2}$, the product over
$j$ is expressed as a sum over $\ell$
\begin{equation}
\frac{\sqrt{\det\Delta_{\mathrm{F}}}}{\sqrt{\det\Delta_{\mathrm{B}}}}=\prod_{\alpha\in\Delta_{+}}\prod_{\ell=1}^{\infty}\left(1\right)^{2\ell+|\alpha(\mathfrak{m})|+1}=\prod_{\alpha\in\Delta_{+}}\left(1\right)^{\sum_{\ell=1}^{\infty}\left(2\ell+|\alpha(\mathfrak{m})|+1\right)}
\end{equation}
The sum is evaluated using Dirichlet regularization 
\begin{equation}
\lim_{\varsigma\to0}(\sum_{\ell=1}^{\infty}(2\ell+|\alpha(\mathfrak{m})|+1)\ell^{-\varsigma})=-\frac{1}{6}(3|\alpha(\mathfrak{m})|+4),
\end{equation}
and the ratio of determinants reduces to
\begin{equation}
\frac{\sqrt{\det\Delta_{\mathrm{F}}}}{\sqrt{\det\Delta_{\mathrm{B}}}}=\prod_{\alpha\in\Delta_{+}}\left(1\right)^{-\frac{|\alpha(\mathfrak{m})|}{2}-\frac{2}{3}}=1.
\end{equation}
For $j=\frac{|\alpha(\mathfrak{m})|}{2}\geq\frac{1}{2},\;\text{\ensuremath{\alpha}(\ensuremath{\mathfrak{m}})\ensuremath{\geq}1},$
the harmonics $Y_{j,j_{3}}^{s},\;Y_{j,j_{3}}^{s+1}$ exist, and the
eigenvalues have multiplicity $|\alpha(\mathfrak{m})|+1$. Removing
the second row and column in (\ref{eq:Obos}) and (\ref{eq:Ofer}),
and noting that $D_{\bar{z}}^{0}Y_{j,j_{3}}^{s}=0$, the contribution
is 
\begin{equation}
\frac{\sqrt{\det\Delta_{\mathrm{F}}}}{\sqrt{\det\Delta_{\mathrm{B}}}}=\prod_{\alpha\in\Delta_{+}}\frac{\left(\frac{i\alpha(\mathfrak{m})^{3}\alpha(u)}{128R^{6}}\right)^{\frac{\left|\alpha(\mathfrak{m})\right|+1}{2}}}{\left(\frac{\alpha(\mathfrak{m})^{3}}{128R^{6}}\right)^{\frac{\left|\alpha(\mathfrak{m})\right|+1}{2}}}
\end{equation}
Similarly, for $j=\frac{|\alpha(\mathfrak{m})|}{2}\geq\frac{1}{2},\;\alpha(\mathfrak{m})\leq-1,$
the harmonics $Y_{j,j_{3}}^{s},\;Y_{j,j_{3}}^{s-1}$ exist, and the
eigenvalues have multiplicity $|\alpha(\mathfrak{m})|+1$. Removing
the first row and column in (\ref{eq:Obos}) and (\ref{eq:Ofer}),
and noting that $D_{z}^{0}Y_{j,j_{3}}^{s}=0$, the contribution is
\begin{equation}
\frac{\sqrt{\det\Delta_{\mathrm{F}}}}{\sqrt{\det\Delta_{\mathrm{B}}}}=\prod_{\alpha\in\Delta_{+}}\frac{\left(-\frac{i\alpha(\mathfrak{m})^{3}\alpha(u)}{128R^{6}}\right)^{\frac{\left|\alpha(\mathfrak{m})\right|+1}{2}}}{\left(-\frac{\alpha(\mathfrak{m})^{3}}{128R^{6}}\right)^{\frac{\left|\alpha(\mathfrak{m})\right|+1}{2}}}
\end{equation}
For $j=\frac{|\alpha(\mathfrak{m})|}{2}-1\geq0,\;\alpha(\mathfrak{m})\geq2,$
the harmonic $Y_{j,j_{3}}^{s+1}$ exists, and the eigenvalues have
multiplicity $|\alpha(\mathfrak{m})|-1$. Removing the second, third,
and fourth rows and columns in (\ref{eq:Obos}) and (\ref{eq:Ofer}),
and noting that $D_{\bar{z}}^{0}Y_{j,j_{3}}^{s+1}=0$, the contribution
is 
\begin{equation}
\frac{\sqrt{\det\Delta_{\mathrm{F}}}}{\sqrt{\det\Delta_{\mathrm{B}}}}=\prod_{\alpha\in\Delta_{+}}\frac{\left(-\frac{\alpha(\mathfrak{m})}{4R^{2}t}\right)^{\frac{\left|\alpha(\mathfrak{m})\right|-1}{2}}}{\left(-\frac{i\alpha(\mathfrak{m})\alpha(u)}{4R^{2}t}\right)^{\frac{\left|\alpha(\mathfrak{m})\right|-1}{2}}}\label{eq:ratio_amgeq2}
\end{equation}
Similarly, for $j=\frac{|\alpha(\mathfrak{m})|}{2}-1\geq0,\;\alpha(\mathfrak{m})\leq-2,$
the harmonic $Y_{j,j_{3}}^{s-1}$ exists, and the eigenvalues have
multiplicity $|\alpha(\mathfrak{m})|-1$. Removing the first, third,
and fourth rows and columns in (\ref{eq:Obos}) and (\ref{eq:Ofer}),
and noting that $D_{z}^{0}Y_{j,j_{3}}^{s-1}=0$, the contribution
is 
\begin{equation}
\frac{\sqrt{\det\Delta_{\mathrm{F}}}}{\sqrt{\det\Delta_{\mathrm{B}}}}=\prod_{\alpha\in\Delta_{+}}\frac{\left(-\frac{\alpha(\mathfrak{m})}{4R^{2}t}\right)^{\frac{\left|\alpha(\mathfrak{m})\right|-1}{2}}}{\left(-\frac{i\alpha(\mathfrak{m})\alpha(u)}{4R^{2}t}\right)^{\frac{\left|\alpha(\mathfrak{m})\right|-1}{2}}}\label{eq:ratio_amleq2}
\end{equation}
Notice that taking $t\to0$ in (\ref{eq:ratio_amgeq2}) and (\ref{eq:ratio_amleq2})
results in inadmissible ``zero-divided-by-zero'' cancellations. 

Collecting the contributions of each mode in the $\alpha(\mathfrak{m})\neq0$
sector, we have 
\begin{equation}
Z_{\text{1-loop}}^{\alpha(\mathfrak{m})\neq0}  =  \prod_{\alpha\in\Delta_{+}}\frac{\left(\frac{i\alpha(\mathfrak{m})^{3}\alpha(u)}{128R^{6}}\right)^{\frac{\left|\alpha(\mathfrak{m})\right|+1}{2}}}{\left(\frac{\alpha(\mathfrak{m})^{3}}{128R^{6}}\right)^{\frac{\left|\alpha(\mathfrak{m})\right|+1}{2}}}\frac{\left(-\frac{i\alpha(\mathfrak{m})^{3}\alpha(u)}{128R^{6}}\right)^{\frac{\left|\alpha(\mathfrak{m})\right|+1}{2}}}{\left(-\frac{\alpha(\mathfrak{m})^{3}}{128R^{6}}\right)^{\frac{\left|\alpha(\mathfrak{m})\right|+1}{2}}}\frac{\left(-\frac{\alpha(\mathfrak{m})}{4R^{2}t}\right)^{\frac{\left|\alpha(\mathfrak{m})\right|-1}{2}}}{\left(-\frac{i\alpha(\mathfrak{m})\alpha(u)}{4R^{2}t}\right)^{\frac{\left|\alpha(\mathfrak{m})\right|-1}{2}}}\frac{\left(-\frac{\alpha(\mathfrak{m})}{4R^{2}t}\right)^{\frac{\left|\alpha(\mathfrak{m})\right|-1}{2}}}{\left(-\frac{i\alpha(\mathfrak{m})\alpha(u)}{4R^{2}t}\right)^{\frac{\left|\alpha(\mathfrak{m})\right|-1}{2}}}.\label{eq:Z1loopContributions}
\end{equation}
After simplifying, (\ref{eq:Z1loopContributions}) reduces to the
expected one-loop determinant of the $A$-model  vector multiplet
in the $\alpha(\mathfrak{m})\neq0$ sector
\begin{eqnarray}
Z_{\text{1-loop}}^{\alpha(\mathfrak{m})\neq0} & = & \left(-1\right)^{\sum_{\alpha\in\Delta_{+}}\alpha(\mathfrak{m})}\prod_{\alpha\in\Delta_{+}}\alpha(u)^{2}\label{eq:Z1loopNonZeroFlux}
\end{eqnarray}

The gauge-fixing is singular in the $j=0,\,\alpha(\mathfrak{m})=0$
sector. In this case, the harmonic $Y_{j,j_{3}}^{s}=Y_{0,j_{3}}^{0}$
exists, and the eigenvalues have multiplicity $1$. We remove the
first and second rows and columns from (\ref{eq:Obos}) and (\ref{eq:Ofer}).
The bosonic operator $\Delta_{\mathrm{B}}^{\prime}$ acting on scalar
harmonics in the basis $\Phi^{(0)}=(\sigma^{\alpha},\tilde{\sigma}_{E}^{\alpha})_{j=\alpha(\mathfrak{m})=0}^{\top}$
is 
\begin{equation}
\left(\begin{array}{cc}
0 & -\frac{i}{4}D_{\mu}^{0}D_{0}^{\mu}\\
-\frac{i}{4}D_{\mu}^{0}D_{0}^{\mu} & \frac{t^{2}}{2}
\end{array}\right)\left(\begin{array}{c}
Y_{0,j_{3}}^{0}\\
Y_{0,j_{3}}^{0}
\end{array}\right)=\left(\begin{array}{cc}
0 & 0\\
0 & \frac{t^{2}}{2}
\end{array}\right)\left(\begin{array}{c}
Y_{0,j_{3}}^{0}\\
Y_{0,j_{3}}^{0}
\end{array}\right),
\end{equation}
where there is a problematic zero along the diagonal. The fermionic
operator $\Delta_{\mathrm{F}}^{\prime}$ acting on scalar harmonics
in the basis $\Psi^{(0)}=(\lambda^{\alpha},\tilde{\lambda}^{\alpha},c^{\alpha},\tilde{c}^{\alpha})_{j=\alpha(\mathfrak{m})=0}^{\top}$
is 
\begin{equation}
\left(\begin{array}{cccc}
t-\frac{i\alpha\left(\sigma\right)}{4} & -\frac{i\alpha\left(\sigma\right)}{4} & 0 & 0\\
-\frac{i\alpha\left(\sigma\right)}{4} & -t-\frac{i\alpha\left(\sigma\right)}{4} & 0 & 0\\
0 & 0 & iD_{\mu}^{0}D_{0}^{\mu} & 0\\
0 & 0 & 0 & -iD_{\mu}^{0}D_{0}^{\mu}
\end{array}\right)\left(\begin{array}{c}
Y_{0,j_{3}}^{0}\\
Y_{0,j_{3}}^{0}\\
Y_{0,j_{3}}^{0}\\
Y_{0,j_{3}}^{0}
\end{array}\right)=\left(\begin{array}{cccc}
t-\frac{i\alpha\left(\sigma\right)}{4} & -\frac{i\alpha\left(\sigma\right)}{4} & 0 & 0\\
-\frac{i\alpha\left(\sigma\right)}{4} & -t-\frac{i\alpha\left(\sigma\right)}{4} & 0 & 0\\
0 & 0 & 0 & 0\\
0 & 0 & 0 & 0
\end{array}\right)\left(\begin{array}{c}
Y_{0,j_{3}}^{0}\\
Y_{0,j_{3}}^{0}\\
Y_{0,j_{3}}^{0}\\
Y_{0,j_{3}}^{0}
\end{array}\right),
\end{equation}
where $\alpha(\sigma)=-\frac{1}{2}\alpha(u)$. The ratio of determinants
$\sqrt{\det\Delta_{\mathrm{F}}^{\prime}}/\sqrt{\det\Delta_{\mathrm{B}}^{\prime}}$
is singular due to $\det\Delta_{\mathrm{B}}^{\prime}=0$, which stems
from the zero along the diagonal of $\Delta_{\mathrm{B}}^{\prime}$.

To treat the singular gauge-fixing, we follow \cite{Blau:1993hj}
and gauge-fix once more by adding a gauge-fixing term to $\bar{\Phi}^{(0)}\Delta_{\mathrm{B}}^{\prime}\Phi^{(0)}+\bar{\Psi}^{(0)}\Delta_{\mathrm{F}}^{\prime}\Psi^{(0)}$.
We define a nilpotent BRST operator $Q_{\mathbf{b}}$ acting as 
\begin{eqnarray}
Q_{\mathbf{b}}\sigma^{\alpha} & = & \alpha\left(\sigma\right)c^{\alpha}\\
Q_{\mathbf{b}}c^{\alpha} & = & 0\\
Q_{\mathbf{b}}\sigma_{0}^{k} & = & 0\\
Q_{\mathbf{b}}\tilde{c}^{\alpha} & = & B^{\alpha}\\
Q_{\mathbf{b}}B^{\alpha} & = & 0
\end{eqnarray}
where $\alpha(\sigma)=\alpha(\mathtt{H}^{k})\sigma_{0}^{k}$, $\sigma_{0}^{k}$
is real, $\sigma^{\alpha}$ is complex, and $B^{\alpha}$ is complex.
Furthermore, we have complex conjugates $\left(\sigma^{\alpha}\right)^{*}=\sigma^{-\alpha}$
and $\left(B^{\alpha}\right)^{*}=B^{-\alpha}$. Next, we define 
\begin{equation}
V_{0}=\frac{1}{2}\tilde{c}^{-\alpha}B^{\alpha}+i\tilde{c}^{-\alpha}\sigma^{\alpha}+i\sigma^{-\alpha}\tilde{c}^{\alpha}.
\end{equation}
Acting on $V_{0}$ with $Q_{\mathbf{b}}$, we obtain a $Q_{\mathbf{b}}$-exact
gauge-fixing term 
\begin{eqnarray}
Q_{\mathbf{b}}V_{0} & = & \frac{1}{2}B^{-\alpha}B^{\alpha}-iB^{-\alpha}\sigma^{\alpha}+i\tilde{c}^{-\alpha}\alpha\left(\sigma\right)c^{\alpha}-i\sigma^{-\alpha}B^{\alpha}+ic^{-\alpha}\alpha\left(\sigma\right)\tilde{c}^{\alpha},
\end{eqnarray}
Completing the square for complex $B^{\alpha}$, the gauge-fixing
term is 
\begin{equation}
Q_{\mathbf{b}}V_{0}=\frac{1}{2}\left|B^{\alpha}-i\sigma^{\alpha}\right|^{2}+\frac{1}{2}\left|\sigma^{\alpha}\right|^{2}+i\tilde{c}^{-\alpha}\alpha\left(\sigma\right)c^{\alpha}+ic^{-\alpha}\alpha\left(\sigma\right)\tilde{c}^{\alpha}
\end{equation}
We integrate out $B^{\alpha}$, resulting in a contribution of 1 and
the elimination of the $\frac{1}{2}\left|B^{\alpha}-i\sigma^{\alpha}\right|^{2}$
term. The remaining gauge-fixing term is 
\begin{equation}
\left(Q_{\mathbf{b}}V_{0}\right)^{\prime}=\frac{1}{2}\left|\sigma^{\alpha}\right|^{2}+i\tilde{c}^{-\alpha}\alpha\left(\sigma\right)c^{\alpha}+ic^{-\alpha}\alpha\left(\sigma\right)\tilde{c}^{\alpha},
\end{equation}
where $\frac{1}{2}\left|\sigma^{\alpha}\right|^{2}$ is a mass term
for the bosonic scalar $\sigma$. The gauge-fixed expression $\bar{\Phi}^{(0)}\Delta_{\mathrm{B}}^{\prime}\Phi^{(0)}+\bar{\Psi}^{(0)}\Delta_{\mathrm{F}}^{\prime}\Psi^{(0)}+(Q_{\mathbf{b}}V_{0})^{\prime}$
is 
\begin{equation}
(\sigma^{-\alpha},\tilde{\sigma}_{E}^{-\alpha})\underset{\Delta_{\mathrm{B}}^{\prime\prime}}{\underbrace{\left(\begin{array}{cc}
\frac{1}{2} & 0\\
0 & \frac{t^{2}}{2}
\end{array}\right)}}\left(\begin{array}{c}
\sigma^{\alpha}\\
\tilde{\sigma}_{E}^{\alpha}
\end{array}\right)+(\tilde{\lambda}^{-\alpha},\lambda^{-\alpha},\tilde{c}^{-\alpha},c^{-\alpha})\underset{\Delta_{\mathrm{F}}^{\prime\prime}}{\underbrace{\left(\begin{array}{cccc}
t-\frac{i\alpha\left(\sigma\right)}{4} & -\frac{i\alpha\left(\sigma\right)}{4} & 0 & 0\\
-\frac{i\alpha\left(\sigma\right)}{4} & -t-\frac{i\alpha\left(\sigma\right)}{4} & 0 & 0\\
0 & 0 & i\alpha\left(\sigma\right) & 0\\
0 & 0 & 0 & i\alpha\left(\sigma\right)
\end{array}\right)}}\left(\begin{array}{c}
\lambda^{\alpha}\\
\tilde{\lambda}^{\alpha}\\
c^{\alpha}\\
\tilde{c}^{\alpha}
\end{array}\right)
\end{equation}
The new bosonic operator $\Delta_{\mathrm{B}}^{\prime\prime}$ no
longer has a zero on the diagonal. Consequently, the bosonic determinant
$\det\Delta_{\mathrm{B}}^{\prime\prime}$ is non-zero and the ratio
of determinants $\sqrt{\det\Delta_{\mathrm{F}}^{\prime\prime}}/\sqrt{\det\Delta_{\mathrm{B}}^{\prime\prime}}$
is non-singular. Evaluating the ratio of determinants, we obtain the
expression for the one-loop contribution in the $j=\alpha(\mathfrak{m})=0$
sector: 
\begin{equation}
Z_{\text{1-loop}}^{\alpha(\mathfrak{m})=0}=\prod_{\alpha\in\Delta_{+}}\left(\frac{t^{2}\alpha\left(\sigma\right)^{2}}{\frac{t^{2}}{4m^{2}}}\right)=\prod_{\alpha\in\Delta_{+}}\alpha\left(u\right)^{2}.\label{eq:Z1loopZeroFlux}
\end{equation}
Notice that (\ref{eq:Z1loopZeroFlux}) agrees with the expression
(\ref{eq:Z1loopNonZeroFlux}) with $\alpha(\mathfrak{m})$ set to
zero. 

\subsection{Result of localization}
To collect the results of localization of the $A$-model  vector
multiplet on the $S^{2}$, we record the one-loop contributions, the
classical contributions, and present the integral formula for $A$-model
correlators. 

The one-loop contribution of the exact action (\ref{eq:QAV(0,t)})
is 
\begin{equation}
Z_{\text{1-loop}}=\left(-1\right)^{\sum_{\alpha\in\Delta_{+}}\alpha\left(\mathfrak{m}\right)}\prod_{\alpha\in\Delta}\alpha(u),\label{eq:oneLoopVplet}
\end{equation}
The classical contribution is obtained by evaluating the on-shell
value of the classical action on the locus configurations (\ref{eq:locus10}).
The exact terms vanish on these configurations, and the non-zero contributions
come from the closed terms. The classical contribution of the closed
twisted chiral superpotential action (\ref{eq:SWa}) is
\begin{equation}
Z_{\text{class}}^{\widetilde{W}}=e^{-\left.S_{\widetilde{W}}\right|_{\text{on-shell}}}=\exp\left(4\pi\text{Tr}\widetilde{W}^{\prime}(u)\cdot\mathfrak{m}\right)\label{eq:ZclW}
\end{equation}
The classical contribution of the closed physical Yang-Mills action
(\ref{eq:ScohYM}) is 
\begin{equation}
Z_{\text{class}}^{\text{YM}}=e^{\left.(\omega+\varepsilon\Theta)\right|_{\text{on-shell}}}=\exp\bigg(\underset{\omega}{\underbrace{2\pi i{\atop }\text{Tr}\;u\cdot\mathfrak{m}}}-\varepsilon\;\underset{\Theta}{\underbrace{\frac{1}{2}\;\text{Tr}\;u^{2}}}\bigg).\label{eq:ZclYM}
\end{equation}
This is the contribution that is necessary to recover the Yang-Mills
partition function from the $A$-model. The $\omega$ part
of (\ref{eq:ZclYM}) can be regarded as (\ref{eq:ZclW}) specialized
to a quadratic superpotential 
\begin{equation}
\left.\left(Z_{\text{class}}^{\widetilde{W}}\right)\right|_{\widetilde{W}(u)=\frac{i}{4}u^{2}}=\exp\left(2\pi i\;\text{Tr}\;u\cdot\mathfrak{m}\right),\label{eq:ZclWquad}
\end{equation}
while the $\varepsilon\,\Theta$ part can be regarded as a closed
gauge-invariant operator 
\begin{equation}
\mathcal{O}(u)=e^{-\frac{\varepsilon}{2}\;\text{Tr}\;u^{2}}.\label{eq:ZclO}
\end{equation}
The classical physical Yang-Mills action (\ref{eq:ZclYM}) can be
expressed in terms of (\ref{eq:ZclW}) and (\ref{eq:ZclO}), which
reads
\begin{equation}
Z_{\text{class}}^{\text{YM}}=\left.\left(Z_{\text{class}}^{\widetilde{W}}\right)\right|_{\widetilde{W}(u)=\frac{i}{4}u^{2}}\mathcal{O}(u).\label{eq:ZclWquadO}
\end{equation}
The correlator formula for the $A$-model vector multiplet
on $S^{2}$ with gauge group $G$ is 
\begin{equation}
\left\langle \mathcal{O}\right\rangle =\frac{\left(-1\right)^{r}}{\left|W\right|}\,\sum_{\mathfrak{m}\in\Lambda_{\text{coch}}^{G}}\,\int_{\mathbb{R}}\text{d}u^{r}\,\mathcal{O}(u)\,e^{4\pi\widetilde{W}^{\prime}\left(u\right)\cdot\mathfrak{m}}\;\left(-1\right)^{\sum_{\alpha\in\Delta_{+}}\alpha\left(\mathfrak{m}\right)}\prod_{\alpha\in\Delta}\alpha\left(u\right),\label{eq:AmodelcorrelatorRevised}
\end{equation}
where $r=\text{rank}G$ and $\mathcal{O}$ is a closed gauge-invariant
operator. The $A$-model  correlator formula (\ref{eq:AmodelcorrelatorRevised})
is nearly identical to the original formula (\ref{eq:intro_AmodelCorrelator}),
except that $u$ is a real modulus that can be integrated along the
real contour $\mathbb{R}$. The partition function formula is obtained
by setting the operator to one in the correlator formula. 

Under the assumption that the localization scheme remains valid at
higher genera, the one-loop determinant of the $A$-model vector
multiplet on a Riemannian manifold $\Sigma_{g}$ of genus $g$ is
\cite{Benini:2016hjo}
\begin{equation}
Z_{\text{1-loop}}=\left(-1\right)^{\sum_{\alpha\in\Delta_{+}}\alpha\left(\mathfrak{m}\right)}\prod_{\alpha\in\Delta}\alpha(u)^{1-g},\label{eq:oneLoopVpletSigma}
\end{equation}
Note that the phase factor can be expressed as $\left(-1\right)^{\sum_{\alpha\in\Delta_{+}}\alpha\left(\mathfrak{m}\right)}=e^{-2\pi i\varrho(\mathfrak{m})}$
using $e^{-i\pi}=-1$ and the Weyl vector $\varrho=\frac{1}{2}\sum_{\alpha\in\Delta_{+}}\alpha$. 

%Let us make a final observation. In \cite{Griguolo:2024ecw}, the
%path integral of pure $\text{YM}_{2}$ on $S^{2}$ with $G=SU(2)$
%was evaluated analytically. Instead of a one-loop determinant, the
%authors derived an integrable distribution on $u$ with singular support.
%It is plausible that the result for higher-rank gauge groups is 
%well-defined products of integrable distributions on $u$, and that
%(\ref{eq:oneLoopVplet}) generalizes this case. 

\section{Recovering two-dimensional Yang-Mills from the $A$-model } \label{sec:3}

To recover the two-dimensional Yang-Mills partition function from
the $A$-model , we will evaluate 
\begin{equation}
\left\langle e^{\omega+\varepsilon\Theta}\right\rangle =\frac{\left(-1\right)^{r}}{\left|W\right|}\int_{\mathbb{R}}\mathrm{d}u^{r}\sum_{\mathfrak{m}\in\Lambda_{\text{coch}}^{G}}e^{2\pi i\left(u,\mathfrak{m}\right)-2\pi i\left\langle \varrho,\mathfrak{m}\right\rangle -\frac{\varepsilon}{2}\left(u,u\right)}\prod_{\alpha\in\Delta}\left\langle \alpha,u\right\rangle ^{1-g}\label{eq:<exp>}
\end{equation}
where $\left(u,\mathfrak{m}\right)=u\cdot\mathfrak{m}$ denotes the
pairing with the scalar product $\left(\cdot,\cdot\right)$ on $\mathfrak{h}$,
and $\left\langle \alpha,u\right\rangle =\alpha\left(u\right)$ denotes
the pairing between dual vector spaces $\left\langle \cdot,\cdot\right\rangle :\mathfrak{h}^{*}\otimes\mathfrak{h}\to\mathbb{R}$.
In particular, (\ref{eq:<exp>}) can either be regarded as the $A$-model
 correlator formula (\ref{eq:AmodelcorrelatorRevised}) with the operator
(\ref{eq:ZclO}) and a quadratic superpotential (\ref{eq:ZclWquad}),
or the expectation value of the classical closed physical Yang-Mills
action (\ref{eq:ZclWquadO}). We will first sum over $\mathfrak{m}$,
then integrate $u$ along the real contour. 

\subsection{Integrals over moduli}

The sum over $\mathfrak{m}\in\Lambda_{\text{coch}}^{G}$ is evaluated
using the Poisson summation formula\footnote{The Poisson summation formula generalized to lattices is $\sum_{x\in L}f(x)=\sum_{y\in L^{*}}\hat{f}(y)$,
where $L$ is a lattice, $L^{*}$ is the integral dual lattice of
$L$, and $\hat{f}$ is the Fourier transform of $f$.}, which yields a periodic delta function on $u$ 
\begin{eqnarray}
\left\langle e^{\omega+\varepsilon\Theta}\right\rangle  & = & \frac{\left(-1\right)^{r}}{\left|W\right|}\int\mathrm{d}u^{r}\left(\sum_{\mathfrak{m}\in\Lambda_{\text{coch}}^{G}}e^{2\pi i\left(u,\mathfrak{m}\right)-2\pi i\left\langle \varrho,\mathfrak{m}\right\rangle }\right)e^{-\frac{\varepsilon}{2}\left(u,u\right)}\prod_{\alpha\in\Delta}\left\langle \alpha,u\right\rangle ^{1-g}\label{eq:sumInitial}\\
 & = & \frac{\left(-1\right)^{r}}{\left|W\right|}\int\mathrm{d}u^{r}\left(\frac{1}{\text{vol}\left(\Lambda_{\text{coch}}^{G}\right)}\sum_{\mu\in\Lambda_{\text{ch}}^{G}}\delta^{(r)}\left(u-\mu-\varrho\right)\right)e^{-\frac{\varepsilon}{2}\left(u,u\right)}\prod_{\alpha\in\Delta}\left\langle \alpha,u\right\rangle ^{1-g}.\label{eq:sumFinal}
\end{eqnarray}
Here, $\delta^{(r)}\left(\cdot\right)$ is the $r$-dimensional delta
function, $\text{vol}\left(\Lambda_{\text{coch}}^{G}\right)$ is the
area of the unit plaquette in $\Lambda_{\text{coch}}^{G}$, and $\Lambda_{\text{ch}}^{G}$
is the integral dual lattice of $\Lambda_{\text{coch}}^{G}$. Notice
that one passes from a sum over the cocharacter lattice $\Lambda_{\text{coch}}^{G}\subset\mathfrak{h}$
in (\ref{eq:sumInitial}) to a sum over the character lattice $\Lambda_{\text{ch}}^{G}\subset\mathfrak{h}^{*}$
in (\ref{eq:sumFinal}). 

Next we integrate $u$ along the real contour. The integral in (\ref{eq:sumFinal})
only receives contributions from the points $u-\mu-\varrho=0$ where
the delta function has support. Consequently, the integral over real
$u$ evaluates to 
\begin{eqnarray}
\left\langle e^{\omega+\varepsilon\Theta}\right\rangle  & = & \frac{\left(-1\right)^{r}}{\left|W\right|\text{vol}\left(\Lambda_{\text{coch}}^{G}\right)}\int_{\mathbb{R}}\mathrm{d}u^{r}\left(\sum_{\mu\in\Lambda_{\text{ch}}^{G}}\delta^{(r)}\left(u-\mu-\varrho\right)e^{-\frac{\varepsilon}{2}\left(u,u\right)}\prod_{\alpha\in\Delta}\left\langle \alpha,u\right\rangle ^{1-g}\right)\label{eq:intInitial}\\
 & = & \frac{\left(-1\right)^{r}}{\left|W\right|\text{vol}\left(\Lambda_{\text{coch}}^{G}\right)}\sum_{\mu\in\Lambda_{\text{ch}}^{G}}e^{-\frac{\varepsilon}{2}\left(\mu+\varrho,\mu+\varrho\right)}\prod_{\alpha\in\Delta}\left(\alpha,\mu+\varrho\right)^{1-g}\label{eq:intFinal}
\end{eqnarray}
Notice that the integrand in (\ref{eq:intInitial}) includes quantities
that are valued in both $\mathfrak{h}$ and $\mathfrak{h}^{*}$, while
the summand in (\ref{eq:intFinal}) only includes quantities that
are valued in $\mathfrak{h}^{*}$. 

\subsection{Eliminating Weyl group invariance}

The Weyl group invariance in (\ref{eq:intFinal}) is eliminated by
using the $W$-action to reflect all $\mathfrak{h}^{*}$-valued quantities
into the fundamental Weyl chamber. Recall that selecting a particular
Cartan subalgebra $\mathfrak{h}$ reduces the gauge group to a subgroup,
and the residual gauge transformations acting on $\mathfrak{h}$ form
the Weyl group $W$ \cite{Kapustin:2005py}. This group is a finite
reflection group that acts on both $\mathfrak{h}$ and its dual $\mathfrak{h}^{*}$
through Weyl reflections $w\in W$. In $\mathfrak{h}^{*}$, these
reflections occur about hyperplanes $\pi_{\alpha}=\{\gamma\subset\mathfrak{h}^{*}\,|\,(\alpha,\gamma)=0\}$
that are orthogonal to roots $\alpha\in\Delta$. The hyperplanes serve
as the boundaries of Weyl chambers. The union of hyperplanes $\Pi=\bigcup_{\alpha\in\Delta}\pi_{\alpha}$
is equivalent to the union of Weyl chamber boundaries, and the complement
$\Pi^{c}=\mathfrak{h}^{*}\backslash\Pi$ is equivalent to the union
of Weyl chamber interiors. In what follows, $C^{0}$ is the closure
of the fundamental Weyl chamber and $C^{0,i}=C^{0}\backslash\Pi$
is the interior of the fundamental Weyl chamber. Moreover, elements
in $C^{0}$ are called dominant, elements in $C^{0,i}$ are called
strictly dominant, and elements in the weight lattice $\Lambda_{\text{wt}}^{\mathfrak{g}}$
in $\mathfrak{h}^{*}$ are called integral. 

To account for the shift by the Weyl vector in (\ref{eq:intFinal}),
we define the $\varrho$-shifted set\footnote{If the gauge group is the universal covering group $G=\widetilde{G}$,
both the character lattice $\Lambda_{\mathrm{ch}}^{\widetilde{G}}$
and the $\varrho$-shifted set $\mathcal{P}$ coincide with the weight
lattice $\Lambda_{\text{wt}}^{\mathfrak{g}}$ because the $\varrho$-shift
maps $\Lambda_{\mathrm{ch}}^{\widetilde{G}}$ to itself. If the gauge
group is any other cover $G=G^{\prime}$, the character lattice is
a sublattice of the weight lattice $\Lambda_{\mathrm{ch}}^{G^{\prime}}\subset\Lambda_{\text{wt}}^{\mathfrak{g}}$
and $\mathcal{P}$ is a infinite set of integral elements. For instance,
if we pick the adjoint cover $G=\text{SO}(3)$, the weight lattice
is the integers $\Lambda_{\text{wt}}^{\mathfrak{su}(2)}\simeq\mathbb{Z}$,
the root lattice is the even integers $\Lambda_{\text{rt}}^{\mathfrak{su}(2)}\simeq\mathbb{Z}_{\text{even}}$,
the character lattice coincides with the root lattice $\Lambda_{\mathrm{ch}}^{\text{SO}(3)}\simeq\Lambda_{\text{rt}}^{\mathfrak{su}(2)}$,
and the $\varrho$-shifted set coincides with the odd integers $\mathcal{P}\simeq\mathbb{Z}_{\text{odd}}$. } 
\begin{equation}
\mathcal{P}=\left\{ \mu+\varrho\,\vert\,\mu\in\Lambda_{\mathrm{ch}}^{G}\right\} ,\label{eq:appB_shiftedSetQ}
\end{equation}
which is an infinite set of integral elements in $\mathfrak{h}^{*}$.
Expressing the sum in (\ref{eq:intFinal}) in terms of (\ref{eq:appB_shiftedSetQ}),
we have 
\begin{eqnarray}
\sum_{\mu\in\Lambda_{\mathrm{ch}}^{G}}e^{-\frac{\varepsilon}{2}\left(\mu+\varrho,\mu+\varrho\right)}\prod_{\alpha\in\Delta}\left(\alpha,\mu+\varrho\right)^{1-g} & = & \sum_{p\in\mathcal{P}}e^{-\frac{\varepsilon}{2}\left(p,p\right)}\prod_{\alpha\in\Delta}\left(\alpha,p\right)^{1-g}.\label{eq:Winv1}
\end{eqnarray}
The $\varrho$-shifted set is decomposed as 
\begin{equation}
\mathcal{P}=\left(\mathcal{P}\backslash\Pi\right)\cup\left(\mathcal{P}\cap\Pi\right),
\end{equation}
where $\mathcal{P}\backslash\Pi$ is valued on Weyl chamber interiors,
and $\mathcal{P}\cap\Pi$ is valued on Weyl chamber boundaries. The
final sum in (\ref{eq:Winv1}) receives contributions from $\mathcal{P}\backslash\Pi$,
but not from $\mathcal{P}\cap\Pi$, because elements on Weyl chamber
boundaries are orthogonal to at least one root and set the Vandermonde
determinant to zero. In particular, we have $\prod_{\alpha\in\Delta}\left(\alpha,p\right)=0$
for all $p\in\mathcal{P}\cap\Pi$. Consequently, (\ref{eq:Winv1})
reduces to a sum over elements valued on Weyl chamber interiors
\begin{eqnarray}
\sum_{p\in\mathcal{P}}e^{-\frac{\varepsilon}{2}\left(p,p\right)}\prod_{\alpha\in\Delta}\left(\alpha,p\right)^{1-g} & = & \sum_{p\in\mathcal{P}\backslash\Pi}e^{-\frac{\varepsilon}{2}\left(p,p\right)}\prod_{\alpha\in\Delta}\left(\alpha,p\right)^{1-g}\label{eq:Winv2}
\end{eqnarray}
The set $\mathcal{P}\backslash\Pi$ is valued on the $W$-orbit through
the subset of strictly dominant elements $\mathcal{P}\cap C^{0,i}$\footnote{Since $\mathcal{P}\backslash\Pi$ is valued on Weyl chamber interiors,
the $W$-action on $\mathcal{P}\backslash\Pi$ is regular, i..e. transitive
and free. Therefore, each $p\in\mathcal{P}\backslash\Pi$ is valued
on the orbit of $W$ through a unique element in the interior of the
fundamental Weyl chamber $\hat{p}\in\mathcal{P}\cap C^{0,i}$. In
particular, for every $p\in\mathcal{P}\backslash\Pi$ there is a unique
$\hat{p}\in\mathcal{P}\cap C^{0,i}$ such that $p\in O_{W}(\hat{p})=\{w\cdot\hat{p}\,\vert\,w\in W\}$.
By extension from elements to infinite sets, we have $\mathcal{P}\backslash\Pi=O_{W}(\mathcal{P}\cap C^{0,i})$. }, and its orbit-decomposition is 
\begin{equation}
\mathcal{P}\backslash\Pi=O_{W}\left(\mathcal{P}\cap C^{0,i}\right)=\left\{ w\cdot p\,\vert\,w\in W,\,p\in\mathcal{P}\cap C^{0,i}\right\} .\label{eq:appB_orbitDecompQregular}
\end{equation}
Under the orbit decomposition (\ref{eq:appB_orbitDecompQregular}),
the sum in (\ref{eq:Winv2}) reduces to
\begin{eqnarray}
\sum_{p\in\mathcal{P}\backslash\Pi}e^{-\frac{\varepsilon}{2}\left(p,p\right)}\prod_{\alpha\in\Delta}\left(\alpha,p\right)^{1-g} & = & \sum_{w\in W}\sum_{p\in\mathcal{P}\cap C^{0,i}}e^{-\frac{\varepsilon}{2}\left(w\cdot p,w\cdot p\right)}\prod_{\alpha\in\Delta}\left(\alpha,w\cdot p\right)^{1-g}\label{eq:Winv3}
\end{eqnarray}
The Weyl reflections acting in the scalar products obey $\left(w\cdot p,w\cdot p\right)=\left(p,p\right)$
and $\left(\alpha,w\cdot p\right)=\left(w^{-1}\cdot\alpha,p\right)$.
Moreover, we have $w^{-1}\alpha=\beta\in\Delta$, because the Weyl
reflections permute the set of roots\footnote{In particular, $W:\Delta\to\Delta,\,\alpha^{\prime}\mapsto\alpha^{\prime\prime}$,
and we have $\alpha^{\prime\prime}=w\cdot\alpha^{\prime}$ for some
$w\in W$. Acting on $\alpha^{\prime\prime}=w\cdot\alpha^{\prime}$
with inverse Weyl reflections, we have $w^{-1}\alpha^{\prime\prime}=w^{-1}w\cdot\alpha^{\prime}=\alpha^{\prime}$.}. Therefore, (\ref{eq:Winv3}) reduces to 
\begin{eqnarray}
\sum_{w\in W}\sum_{p\in\mathcal{P}\cap C^{0,i}}e^{-\frac{\varepsilon}{2}\left(w\cdot p,w\cdot p\right)}\prod_{\alpha\in\Delta}\left(\alpha,w\cdot p\right)^{1-g} & = & \sum_{w\in W}\sum_{p\in\mathcal{P}\cap C^{0,i}}e^{-\frac{\varepsilon}{2}\left(p,p\right)}\prod_{\alpha\in\Delta}\left(w^{-1}\alpha,p\right)^{1-g}.\\
 & = & \sum_{w\in W}\sum_{p\in\mathcal{P}\cap C^{0,i}}e^{-\frac{\varepsilon}{2}\left(p,p\right)}\prod_{\beta\in\Delta}\left(\beta,p\right)^{1-g}
\end{eqnarray}
As the summand is independent of Weyl reflections, the sum over $w\in W$
evaluates to
\begin{eqnarray}
\sum_{w\in W}\sum_{p\in\mathcal{P}\cap C^{0,i}}e^{-\frac{\varepsilon}{2}\left(p,p\right)}\prod_{\alpha\in\Delta}\left(\alpha,p\right)^{1-g} & = & \left|W\right|\sum_{p\in\mathcal{P}\cap C^{0,i}}e^{-\frac{\varepsilon}{2}\left(p,p\right)}\prod_{\alpha\in\Delta}\left(\alpha,p\right)^{1-g}\label{eq:Winv4}
\end{eqnarray}
where $\left|W\right|$ is the order of the Weyl group. In particular,
the summand in (\ref{eq:Winv3}) is invariant under the action of
the Weyl group, and the factor of $\left|W\right|$ cancels with $\left|W\right|^{-1}$
in (\ref{eq:intFinal}). 

The set of roots in (\ref{eq:Winv4}) is decomposed into positive
and negative subsets $\Delta=\Delta_{+}\cup\Delta_{-}$, such that
for each positive root $\alpha\in\Delta_{+}$ there is an associated
negative root $-\alpha\in\Delta_{-}$. Under this decomposition, the
product over roots simplifies as 
\begin{eqnarray}
\prod_{\alpha\in\Delta}\left(\alpha,p\right)^{1-g} & = & \left(\prod_{\alpha\in\Delta_{+}}\left(\alpha,p\right)\prod_{\alpha\in\Delta_{-}}\left(\alpha,p\right)\right)^{1-g}\\
 & = & \left(\prod_{\alpha\in\Delta_{+}}\left(\alpha,p\right)\prod_{\alpha\in\Delta_{+}}\left(-\alpha,p\right)\right)^{1-g}\\
 & = & \left(\left(-1\right)^{\left|\Delta_{+}\right|}\right)^{1-g}\prod_{\alpha\in\Delta_{+}}\left(\alpha,p\right)^{2-2g},\label{eq:appB_prodRootsPos}
\end{eqnarray}
where the last equality is due to $\left(\alpha,p\right)>0$ for $\alpha\in\Delta_{+}$
and $p\in\mathcal{P}\cap C^{0,i}$. Replacing the product over roots
in (\ref{eq:Winv4}) with (\ref{eq:appB_prodRootsPos}), the sum reads
\begin{equation}
\sum_{p\in\mathcal{P}\cap C^{0,i}}e^{-\frac{\varepsilon}{2}\left(p,p\right)}\prod_{\alpha\in\Delta}\left(\alpha,p\right)^{1-g}=\left(\left(-1\right)^{\left|\Delta_{+}\right|}\right)^{1-g}\sum_{p\in\mathcal{P}\cap C^{0,i}}e^{-\frac{\varepsilon}{2}\left(p,p\right)}\prod_{\alpha\in\Delta_{+}}\left(\alpha,p\right)^{2-2g}\label{eq:Winv5}
\end{equation}
The sets $\mathcal{P}\cap C^{0,i}$ and $\Lambda_{\text{ch}}^{G}\cap C^{0}$
are in one-to-one correspondence because the $\varrho$-shift is a
bijective map between the dominant and strictly dominant integral
elements. 
Consequently, (\ref{eq:Winv5}) can be expressed as 
\begin{eqnarray}
\sum_{p\in\mathcal{P}\cap C^{0,i}}e^{-\frac{\varepsilon}{2}\left(p,p\right)}\prod_{\alpha\in\Delta_{+}}\left(\alpha,p\right)^{2-2g} & = & \sum_{\mu\in\Lambda_{\text{ch}}^{G}\cap C^{0}}e^{-\frac{\varepsilon}{2}\left(\mu+\varrho,\mu+\varrho\right)}\prod_{\alpha\in\Delta_{+}}\left(\alpha,\mu+\varrho\right)^{2-2g}.\label{eq:Winv6}
\end{eqnarray}
As all quantities in (\ref{eq:Winv6}) are valued in the fundamental
Weyl chamber, we have eliminated the Weyl group invariance. To stress
this point, we express the set being summed over as $\Lambda_{\text{ch}}^{G}/W=\Lambda_{\text{ch}}^{G}\cap C^{0}$.
Collecting the factors, the expectation value of the $A$-model
 correlator in (\ref{eq:intFinal}) now reads 
\begin{equation}
\left\langle e^{\omega+\varepsilon\Theta}\right\rangle =\frac{\left(-1\right)^{r}\left(\left(-1\right)^{\left|\Delta_{+}\right|}\right)^{1-g}}{\text{vol}\left(\Lambda_{\text{coch}}^{G}\right)}\sum_{\mu\in\Lambda_{\text{ch}}^{G}/W}e^{-\frac{\varepsilon}{2}\left(\mu+\varrho,\mu+\varrho\right)}\prod_{\alpha\in\Delta_{+}}\left(\alpha,\mu+\varrho\right)^{2-2g}\label{eq:Winv7}
\end{equation}

\subsection{Final recovery}

To recover the partition function of pure two-dimensional Yang-Mills
theory from the vev of the $A$-model  operator (\ref{eq:Winv7}),
we use the fact that the dominant elements of the character lattice
are in one-to-one correspondence with irreducible representations
of the gauge group, i.e. $\Lambda_{\text{ch}}^{G}/W\simeq\text{irreps}(G)$
\cite{simon1996representations}. Given an element $\mu\in\Lambda_{\text{ch}}^{G}/W$,
the dimension and quadratic Casimir of an element $R_{\mu}\in\text{irreps}(G)$
are given by 
\begin{equation}
\dim\left(R_{\mu}\right)=\frac{\prod_{\alpha\in\Delta_{+}}\left(\alpha,\mu+\varrho\right)}{\prod_{\alpha\in\Delta_{+}}\left(\alpha,\varrho\right)},\qquad C_{2}\left(R_{\mu}\right)=\left(\mu+\varrho,\mu+\varrho\right)-\left(\varrho,\varrho\right).\label{eq:Winv8}
\end{equation}
The first formula is the Weyl dimension formula, and the second is
a similar formula for the quadratic Casimir. For a reference, see
equations 4.23 and 4.24 in \cite{Blau:1996ct}. In view of (\ref{eq:Winv8}),
we can express (\ref{eq:Winv7}) as
\begin{eqnarray}
\left\langle e^{\omega+\varepsilon\Theta}\right\rangle  & = & \frac{\left(-1\right)^{r}\left(\left(-1\right)^{\left|\Delta_{+}\right|}\right)^{1-g}}{\text{vol}\left(\Lambda_{\text{coch}}^{G}\right)}\sum_{R_{\mu}\in\text{irreps}(G)}e^{-\frac{\varepsilon}{2}\left(C_{2}\left(R_{\mu}\right)+\left(\varrho,\varrho\right)\right)}\left(\dim\left(R_{\mu}\right)\prod_{\alpha\in\Delta_{+}}\left(\alpha,\varrho\right)\right)^{2-2g}\\
 & = & \frac{\left(-1\right)^{r}\left(\left(-1\right)^{\left|\Delta_{+}\right|}\right)^{1-g}\left(\prod_{\alpha\in\Delta_{+}}\left(\alpha,\varrho\right)\right)^{2-2g}}{\text{vol}\left(\Lambda_{\text{coch}}^{G}\right)}\sum_{R_{\mu}\in\text{irreps}(G)}\frac{e^{-\frac{\varepsilon}{2}\left(C_{2}\left(R_{\mu}\right)+\left(\varrho,\varrho\right)\right)}}{\dim\left(R_{\mu}\right)^{2g-2}}\label{eq:Winv9}
\end{eqnarray}
 The final expression is the partition function of pure two-dimensional
Yang-Mills theory up to renormalization scheme dependent constants.

\section{Discussion } \label{sec:4}

In this paper, we applied localization techniques to the $A$-model
 vector multiplet on $S^{2}$ with gauge group $G$, derived a formula
(\ref{eq:AmodelcorrelatorRevised}) in which partition functions and
correlators of the $A$-model  are described by integrals along
a real contour, and used the correlator formula to recover the partition
function of two-dimensional Yang-Mills theory (\ref{eq:Winv9}) from
the vacuum expectation value of an $A$-model  operator (\ref{eq:<exp>}).
We demonstrated that non-abelian localization techniques 
successfully retain the necessary contributions of unstable
instantons to the path integral of the $A$-model  on $S^{2}$,
and provide an alternative to the supersymmetric localization techniques
employing JK residues used in \cite{Benini:2015noa,Benini:2016hjo,Closset:2015rna}.
This work is a step toward the potential reconciliation of the discrepancy
between \emph{i)} conventional supersymmetric localization techniques
established in \cite{Pestun:2007rz}; \emph{ii)} non-abelian localization
techniques established in \cite{Witten:1992xu}; and \emph{iii)} supersymmetric
localization techniques employing JK residues established in \cite{Benini:2013nda}.
It would be interesting to develop criteria on the target space or
supersymmetry algebra that further clarify the scope of these distinct
localization techniques. Ideally, the criteria would aid in identifying
appropriate localizing terms and problematic field configurations
that spoil localization. 

A natural extension of this work is the incorporation of chiral multiplets in the
localization of the $A$-model  on $S^{2}$, and the derivation
of formulae for partition functions and correlators described by integrals
along a real contour. The one-loop determinant of a chiral multiplet
could be computed in at least two ways. Firstly, it could be calculated
mode-by-mode using monopole spherical harmonics on $S^{2}$. Alternatively,
it could potentially be derived by reversing the Higgs-mechanism argument
in \cite{Benini:2015noa,Benini:2016hjo,Closset:2015rna}. This argument
relies on the understanding that the contributions of a supersymmetric
W-boson are equivalent to those of a chiral multiplet with an R-charge
of two, and their one-loop determinants are inversely related. These integral
formulae would have direct applications to the genus-zero computation
of \emph{i)} $A$-model  topological amplitudes and Gromov-Witten
(GW) invariants \cite{Jockers:2012dk,Gomis:2012wy};\emph{ }and \emph{ii)} the
volume of the vortex moduli space \cite{Ohta:2019odi}. Another promising
application is testing non-perturbative dualities \cite{Closset:2017vvl}.

A second potential generalization involves the localization of $A$-model
vector multiplet theories on the torus $T^{2}$, extending the genus
zero derivation in section 2 to the genus one case. An important objective
would be to clarify the role of unstable instantons in the localization
of $A$-model theories on $T^{2}$. Although Atiyah and Bott classified
these unstable instantons in \cite{Atiyah:1982fa}, and Griguolo demonstrated
that they contribute to pure $\text{YM}_{2}$ theories on $T^{2}$ in \cite{Griguolo:1998kq},
it is expected that they do not contribute to $A$-model theories
on $T^{2}$. Localizing $A$-model vector multiplet theories on $T^{2}$,
rather than $S^{2}$, introduces several novelties, including non-trivial
flat connections, two half-integer gauge fluxes, and the presence
of fermionic zero modes. The localization computation is expected
to involve solving the $\text{YM}_{2}$ equations on $T^{2}$ for a specific
gauge group, e.g. $G=SU(N)$. The evaluation of the one-loop contribution
will likely require Fourier expansions and involve a fluctuation operator
that is a supermatrix with odd entries, making the computation of
the fluctuation superdeterminant more involved than in the genus zero
case. This analysis could provide insight into the somewhat opaque
formula for the one-loop determinant of the $A$-model vector multiplet
on a genus $g$ Riemannian manifold (\ref{eq:oneLoopVpletSigma}), which evaluates to one
at $g=1$. In contrast to the examples of localization of $\mathcal{N}=(2,2)$
theories on $T^{2}$ in \cite{Benini:2016hjo,Benini:2013nda,Benini:2013xpa}, the derivation would not rely on the JK
residue prescription. As a potential application, one could compute
an $A$-model observable: the partition function of pure $\text{YM}_{2}$ on
$T^{2}$ for $G=SU(N)$, expressed as a sum over contributions localized
around classical solutions of the $\text{YM}_{2}$ equations. This would
generalize the partition function of pure $\text{YM}_{2}$ on $T^{2}$ for
$G=U(N)$ that was derived in \cite{Griguolo:1998kq}.

Let us make a final observation. In \cite{Griguolo:2024ecw},
the partition function of pure $\text{YM}_{2}$ on a genus $g$
Riemannian manifold $\Sigma_{g}$ for $G=SU(2)$ was computed exactly
as a perturbative expansion around unstable instanton sectors. To
achieve this, the authors used a stationary phase version of supersymmetric
localization to evaluate path integrals of a cohomological $\text{YM}_{2}$
theory on $\Sigma_{g}$. This particular cohomological theory involved
the fields of the basic multiplet $(A_{\mu},\psi_{\mu},\phi)$, rather
than the full standard multiplet 
appearing in equations (\ref{eq:dictStaVec}) and (\ref{eq:staplet}). Surprisingly, the localization
computation resulted in a one-loop contribution that was an integrable
distribution on $u$ with singular support. For higher-rank gauge groups, 
it is plausible that the evaluation of the one-loop contribution results in well-defined products of integrable 
distributions on $u$, and that the determinant (\ref{eq:oneLoopVplet}) generalizes this case. 
It would be interesting to extend this stationary phase 
version of localization to an $A$-model setting
and evaluate a one-loop contribution as a distribution on $u$. However,
we expect this approach to be most feasible for $A$-model theories
involving both vector and chiral multiplets. Furthermore, we expect 
that the unstable instanton expansion of the pure $\text{YM}_{2}$ partition function, 
as derived in \cite{Griguolo:2024ecw}, could be recovered from the  formula (\ref{eq:<exp>}) 
by first integrating over $u$ for fixed $\mathfrak{m}$, then expanding in small $\varepsilon$. 
This  derivation, however, is left for future investigation.

\acknowledgments
I would like to thank Luca Griguolo, Itamar Yaakov, and Can Kozçaz for numerous useful discussions and comments on the draft. I would also like to thank Charlotte Kristjansen and the other faculty members at the Niels Bohr Institute at Copenhagen University for their hospitality. This work has received funding from the European Union’s Horizon 2020 research and innovation program under the Marie Skłodowska-Curie grant agreement No. 813942 (ITN EuroPLEx). 

\appendix
\section{$A$-twisted $\mathcal{N}=(2,2)$ vector multiplet} \label{appendix}

Here we provide some details regarding the $A$-twist of the $\mathcal{N}=(2,2)$
vector multiplet on $\mathbb{R}^{2}$. After expanding the contracted
spinors in (\ref{eq:susyR2}) the variation
\[
\delta=\frac{1}{\sqrt{2}}\left(\epsilon_{+}Q_{-}-\epsilon_{-}Q_{+}+\tilde{\epsilon}_{+}\widetilde{Q}_{-}-\tilde{\epsilon}_{-}\widetilde{Q}_{+}\right)
\]
acts on the fields of the $\mathcal{N}=(2,2)$ vector multiplet on
$\mathbb{R}^{2}$ as 

\begin{equation}
\begin{array}{l}
\delta A_{1}=-\frac{i}{\sqrt{2}}\left(\epsilon_{-}\tilde{\lambda}_{-}+\tilde{\epsilon}_{-}\lambda_{-}-\epsilon_{+}\tilde{\lambda}_{+}-\tilde{\epsilon}_{+}\lambda_{+}\right)\\
\delta A_{2}=-\frac{1}{\sqrt{2}}\left(\epsilon_{-}\tilde{\lambda}_{-}+\tilde{\epsilon}_{-}\lambda_{-}+\epsilon_{+}\tilde{\lambda}_{+}+\tilde{\epsilon}_{+}\lambda_{+}\right)\\
\delta(\sigma+\tilde{\sigma})=\sqrt{2}\left(\epsilon_{-}\tilde{\lambda}_{+}-\epsilon_{+}\tilde{\lambda}_{-}-\tilde{\epsilon}_{-}\lambda_{+}+\tilde{\epsilon}_{+}\lambda_{-}\right)\\
\delta(\sigma-\tilde{\sigma})=-\sqrt{2}\left(\epsilon_{-}\tilde{\lambda}_{+}+\epsilon_{+}\tilde{\lambda}_{-}+\tilde{\epsilon}_{-}\lambda_{+}+\tilde{\epsilon}_{+}\lambda_{-}\right)\\
\delta\sigma=-\sqrt{2}\left(\epsilon_{+}\tilde{\lambda}_{-}+\tilde{\epsilon}_{-}\lambda_{+}\right)\\
\delta\tilde{\sigma}=\sqrt{2}\left(\epsilon_{-}\tilde{\lambda}_{+}+\tilde{\epsilon}_{+}\lambda_{-}\right)\\
\delta D=\frac{i\left(D_{1}\epsilon_{-}\tilde{\lambda}_{-}-D_{1}\tilde{\epsilon}_{-}\lambda_{-}-iD_{2}\epsilon_{-}\tilde{\lambda}_{-}+iD_{2}\tilde{\epsilon}_{-}\lambda_{-}-D_{1}\epsilon_{+}\tilde{\lambda}_{+}+D_{1}\tilde{\epsilon}_{+}\lambda_{+}-iD_{2}\epsilon_{+}\tilde{\lambda}_{+}+iD_{2}\tilde{\epsilon}_{+}\lambda_{+}\right)}{\sqrt{2}}\\
\qquad+\frac{i\left(\epsilon_{-}\left[\sigma,\tilde{\lambda}_{+}\right]-\epsilon_{+}\left[\tilde{\sigma},\tilde{\lambda}_{-}\right]+\tilde{\epsilon}_{-}\left[\tilde{\sigma},\lambda_{+}\right]-\tilde{\epsilon}_{+}\left[\sigma,\lambda_{-}\right]\right)}{\sqrt{2}}\\
\delta\lambda_{+}=\frac{i\epsilon_{+}\left[\sigma,\tilde{\sigma}\right]+2\epsilon_{+}D+4i\epsilon_{-}\left(\frac{1}{2}\left(D_{1}-iD_{2}\right)\right)\sigma-2i\epsilon_{+}F_{12}}{2\sqrt{2}}\\
\delta\lambda_{-}=\frac{2\epsilon_{-}D+i\left(-\epsilon_{-}\left[\sigma,\tilde{\sigma}\right]+4\epsilon_{+}\left(\frac{1}{2}\left(D_{1}+iD_{2}\right)\right)\tilde{\sigma}+2\epsilon_{-}F_{12}\right)}{2\sqrt{2}}\\
\delta\tilde{\lambda}_{+}=-\frac{2\tilde{\epsilon}_{+}D+i\left(\tilde{\epsilon}_{+}\left[\sigma,\tilde{\sigma}\right]+4\tilde{\epsilon}_{-}\left(\frac{1}{2}\left(D_{1}-iD_{2}\right)\right)\tilde{\sigma}+2\tilde{\epsilon}_{+}F_{12}\right)}{2\sqrt{2}}\\
\delta\tilde{\lambda}_{-}=\frac{-2\tilde{\epsilon}_{-}D+i\left(\tilde{\epsilon}_{-}\left[\sigma,\tilde{\sigma}\right]-4\tilde{\epsilon}_{+}\left(\frac{1}{2}\left(D_{1}+iD_{2}\right)\right)\sigma+2\tilde{\epsilon}_{-}F_{12}\right)}{2\sqrt{2}}\\
\delta F_{12}=-\frac{-i\left(D_{2}\epsilon_{-}\tilde{\lambda}_{-}+D_{2}\tilde{\epsilon}_{-}\lambda_{-}-D_{2}\epsilon_{+}\tilde{\lambda}_{+}-D_{2}\tilde{\epsilon}_{+}\lambda_{+}\right)+D_{1}\epsilon_{-}\tilde{\lambda}_{-}+D_{1}\tilde{\epsilon}_{-}\lambda_{-}+D_{1}\epsilon_{+}\tilde{\lambda}_{+}+D_{1}\tilde{\epsilon}_{+}\lambda_{+}}{\sqrt{2}}
\end{array}
\end{equation}
To obtain the transformations of the $A$-model vector multiplet
on $S^{2}$, we set $(\epsilon_{+},\epsilon_{-},\tilde{\epsilon}_{+},\tilde{\epsilon}_{-})$
to $(1,0,0,-1)$ and rescale $\delta\to\sqrt{2}\delta$ in the above
transformations. The supercharges $Q:=Q_{+}$, $\widetilde{Q}:=\widetilde{Q}_{-}$,
and $Q_{A}=Q+\widetilde{Q}$ act on the fields of $A$-model
  vector multiplet as 
\begin{equation}
\begin{array}{lcll}
QA_{z}=0 &  & \widetilde{Q}A_{z}=i\lambda_{z} & Q_{A}A_{z}=i\lambda_{z}\\
QA_{\bar{z}}=i\lambda_{\bar{z}} &  & \widetilde{Q}A_{\bar{z}}=0 & Q_{A}A_{\bar{z}}=i\lambda_{\bar{z}}\\
Q\sigma=0 &  & \widetilde{Q}\sigma=0 & Q_{A}\sigma=0\\
Q\tilde{\sigma}=-2\tilde{\lambda} &  & \widetilde{Q}\tilde{\sigma}=2\lambda & Q_{A}\tilde{\sigma}=2\lambda-2\tilde{\lambda}\\
QD=-2iD_{z}\lambda_{\bar{z}}-i\left[\sigma,\tilde{\lambda}\right] &  & \widetilde{Q}D=2iD_{\bar{z}}\lambda_{z}-i\left[\sigma,\lambda\right] & Q_{A}D=-2iD_{z}\lambda_{\bar{z}}+2iD_{\bar{z}}\lambda_{z}-i[\sigma,\lambda+\tilde{\lambda}]\\
Q\lambda_{z}=-2iD_{z}\sigma &  & \widetilde{Q}\lambda_{z}=0 & Q_{A}\lambda_{z}=-2iD_{z}\sigma\\
Q\lambda=\frac{i}{2}\left[\sigma,\tilde{\sigma}\right]-D-i\star F &  & \widetilde{Q}\lambda=0 & Q_{A}\lambda=-D-i\star F+\frac{i}{2}\left[\sigma,\tilde{\sigma}\right]\\
Q\tilde{\lambda}=0 &  & \widetilde{Q}\tilde{\lambda}=-\frac{i}{2}\left[\sigma,\tilde{\sigma}\right]-D-i\star F & Q_{A}\tilde{\lambda}=-D-i\star F-\frac{i}{2}\left[\sigma,\tilde{\sigma}\right]\\
Q\lambda_{\bar{z}}=0 &  & \widetilde{Q}\lambda_{\bar{z}}=-2iD_{\bar{z}}\sigma & Q_{A}\lambda_{\bar{z}}=-2iD_{\bar{z}}\sigma\\
Q\star F=2D_{z}\lambda_{\bar{z}} &  & \widetilde{Q}\star F=-2D_{\bar{z}}\lambda_{z} & Q_{A}\star F=2D_{z}\lambda_{\bar{z}}-2D_{\bar{z}}\lambda_{z}
\end{array}
\end{equation}
where $\star F=F_{12}=-2iF_{z\bar{z}}$ and $\star$ is the Hodge
operator on the $S^{2}$. 

%\appendix
%\section{$A$-twisted $\mathcal{N}=(2,2)$ vector multiplet} \label{appendix}

%\paragraph{Note added.} This is also a good position for notes added
%after the paper has been written.

% Bibliography

%% [A] Recommended: using JHEP.bst file
 \bibliographystyle{JHEP}
% \bibliography{researchStatement_2024}

\providecommand{\href}[2]{#2}\begingroup\raggedright\endgroup

%% or
%% [B] Manual formatting (see below)
%% (i) We suggest to always provide author, title and journal data or doi:
%% in short all the informations that clearly identify a document.
%% (ii) please avoid comments such as "For a review'', "For some examples",
%% "and references therein" or move them in the text. In general, please leave only references in the bibliography and move all
%% accessory text in footnotes.
%% (iii) Also, please have only one work for each \bibitem.

%\printbibliography
%\bibliographystyle{ytphys}
%\baselineskip=0.99\baselineskip
%\bibliography{biblio}

\begin{thebibliography}{10}

\bibitem{Pestun:2007rz}
V.~Pestun, \emph{{Localization of gauge theory on a four-sphere and supersymmetric Wilson loops}}, \href{https://doi.org/10.1007/s00220-012-1485-0}{\emph{Commun. Math. Phys.} {\bfseries 313} (2012) 71} [\href{https://arxiv.org/abs/0712.2824}{{\ttfamily 0712.2824}}].

\bibitem{Benini:2015noa}
F.~Benini and A.~Zaffaroni, \emph{{A topologically twisted index for three-dimensional supersymmetric theories}}, \href{https://doi.org/10.1007/JHEP07(2015)127}{\emph{JHEP} {\bfseries 07} (2015) 127} [\href{https://arxiv.org/abs/1504.03698}{{\ttfamily 1504.03698}}].

\bibitem{Benini:2016hjo}
F.~Benini and A.~Zaffaroni, \emph{{Supersymmetric partition functions on Riemann surfaces}}, {\emph{Proc. Symp. Pure Math.} {\bfseries 96} (2017) 13} [\href{https://arxiv.org/abs/1605.06120}{{\ttfamily 1605.06120}}].

\bibitem{Closset:2015rna}
C.~Closset, S.~Cremonesi and D.S.~Park, \emph{{The equivariant A-twist and gauged linear sigma models on the two-sphere}}, \href{https://doi.org/10.1007/JHEP06(2015)076}{\emph{JHEP} {\bfseries 06} (2015) 076} [\href{https://arxiv.org/abs/1504.06308}{{\ttfamily 1504.06308}}].

\bibitem{Benini:2012ui}
F.~Benini and S.~Cremonesi, \emph{{Partition Functions of ${\mathcal{N}=(2,2)}$ Gauge Theories on S$^{2}$ and Vortices}}, \href{https://doi.org/10.1007/s00220-014-2112-z}{\emph{Commun. Math. Phys.} {\bfseries 334} (2015) 1483} [\href{https://arxiv.org/abs/1206.2356}{{\ttfamily 1206.2356}}].

\bibitem{Doroud:2012xw}
N.~Doroud, J.~Gomis, B.~Le~Floch and S.~Lee, \emph{{Exact Results in D=2 Supersymmetric Gauge Theories}}, \href{https://doi.org/10.1007/JHEP05(2013)093}{\emph{JHEP} {\bfseries 05} (2013) 093} [\href{https://arxiv.org/abs/1206.2606}{{\ttfamily 1206.2606}}].

\bibitem{Doroud:2013pka}
N.~Doroud and J.~Gomis, \emph{{Gauge theory dynamics and K\"ahler potential for Calabi-Yau complex moduli}}, \href{https://doi.org/10.1007/JHEP12(2013)099}{\emph{JHEP} {\bfseries 12} (2013) 099} [\href{https://arxiv.org/abs/1309.2305}{{\ttfamily 1309.2305}}].

\bibitem{Gomis:2012wy}
J.~Gomis and S.~Lee, \emph{{Exact Kahler Potential from Gauge Theory and Mirror Symmetry}}, \href{https://doi.org/10.1007/JHEP04(2013)019}{\emph{JHEP} {\bfseries 04} (2013) 019} [\href{https://arxiv.org/abs/1210.6022}{{\ttfamily 1210.6022}}].

\bibitem{Melnikov:2006kb}
I.V.~Melnikov and M.R.~Plesser, \emph{{A-model correlators from the Coulomb branch}}, \href{https://doi.org/10.1088/1126-6708/2006/02/044}{\emph{JHEP} {\bfseries 02} (2006) 044}.

\bibitem{Benini:2015eyy}
F.~Benini, K.~Hristov and A.~Zaffaroni, \emph{{Black hole microstates in AdS$_{4}$ from supersymmetric localization}}, \href{https://doi.org/10.1007/JHEP05(2016)054}{\emph{JHEP} {\bfseries 05} (2016) 054} [\href{https://arxiv.org/abs/1511.04085}{{\ttfamily 1511.04085}}].

\bibitem{Closset:2017vvl}
C.~Closset, N.~Mekareeya and D.S.~Park, \emph{{A-twisted correlators and Hori dualities}}, \href{https://doi.org/10.1007/JHEP08(2017)101}{\emph{JHEP} {\bfseries 08} (2017) 101} [\href{https://arxiv.org/abs/1705.04137}{{\ttfamily 1705.04137}}].

\bibitem{Witten:1992xu}
E.~Witten, \emph{{Two-dimensional gauge theories revisited}}, \href{https://doi.org/10.1016/0393-0440(92)90034-X}{\emph{J. Geom. Phys.} {\bfseries 9} (1992) 303} [\href{https://arxiv.org/abs/hep-th/9204083}{{\ttfamily hep-th/9204083}}].

\bibitem{Ohta:2019odi}
K.~Ohta and N.~Sakai, \emph{{Higgs and Coulomb Branch Descriptions of the Volume of the Vortex Moduli Space}}, \href{https://doi.org/10.1093/ptep/ptz016}{\emph{PTEP} {\bfseries 2019} (2019) 043B01} [\href{https://arxiv.org/abs/1811.03824}{{\ttfamily 1811.03824}}].

\bibitem{Benini:2013nda}
F.~Benini, R.~Eager, K.~Hori and Y.~Tachikawa, \emph{{Elliptic genera of two-dimensional N=2 gauge theories with rank-one gauge groups}}, \href{https://doi.org/10.1007/s11005-013-0673-y}{\emph{Lett. Math. Phys.} {\bfseries 104} (2014) 465} [\href{https://arxiv.org/abs/1305.0533}{{\ttfamily 1305.0533}}].

\bibitem{Kapustin:2009kz}
A.~Kapustin, B.~Willett and I.~Yaakov, \emph{{Exact Results for Wilson Loops in Superconformal Chern-Simons Theories with Matter}}, \href{https://doi.org/10.1007/JHEP03(2010)089}{\emph{JHEP} {\bfseries 03} (2010) 089} [\href{https://arxiv.org/abs/0909.4559}{{\ttfamily 0909.4559}}].

\bibitem{Blau:1993hj}
M.~Blau and G.~Thompson, \emph{{Lectures on 2-d gauge theories: Topological aspects and path integral techniques}},  in \emph{{Summer School in High-energy Physics and Cosmology (Includes Workshop on Strings, Gravity, and Related Topics 29-30 Jul 1993)}}, pp.~0175--244, 10, 1993 [\href{https://arxiv.org/abs/hep-th/9310144}{{\ttfamily hep-th/9310144}}].

\bibitem{Benini:2016qnm}
F.~Benini and B.~Le~Floch, \emph{{Supersymmetric localization in two dimensions}}, \href{https://doi.org/10.1088/1751-8121/aa77bb}{\emph{J. Phys. A} {\bfseries 50} (2017) 443003} [\href{https://arxiv.org/abs/1608.02955}{{\ttfamily 1608.02955}}].

\bibitem{Witten:1988xj}
E.~Witten, \emph{{Topological Sigma Models}}, \href{https://doi.org/10.1007/BF01466725}{\emph{Commun. Math. Phys.} {\bfseries 118} (1988) 411}.

\bibitem{Closset:2014pda}
C.~Closset and S.~Cremonesi, \emph{{Comments on $ \mathcal{N} $ = (2, 2) supersymmetry on two-manifolds}}, \href{https://doi.org/10.1007/JHEP07(2014)075}{\emph{JHEP} {\bfseries 07} (2014) 075} [\href{https://arxiv.org/abs/1404.2636}{{\ttfamily 1404.2636}}].

\bibitem{Atiyah:1982fa}
M.F.~Atiyah and R.~Bott, \emph{{The Yang-Mills equations over Riemann surfaces}}, {\emph{Phil. Trans. Roy. Soc. Lond. A} {\bfseries 308} (1982) 523}.

\bibitem{Deligne:1999qp}
P.~Deligne, P.~Etingof, D.S.~Freed, L.C.~Jeffrey, D.~Kazhdan, J.W.~Morgan et~al., eds., \emph{{Quantum fields and strings: A course for mathematicians. Vol. 1, 2}} (1999).

\bibitem{Kapustin:2005py}
A.~Kapustin, \emph{{Wilson-'t Hooft operators in four-dimensional gauge theories and S-duality}}, \href{https://doi.org/10.1103/PhysRevD.74.025005}{\emph{Phys. Rev. D} {\bfseries 74} (2006) 025005} [\href{https://arxiv.org/abs/hep-th/0501015}{{\ttfamily hep-th/0501015}}].

\bibitem{simon1996representations}
B.~Simon, \emph{Representations of Finite and Compact Groups}, Graduate studies in mathematics, American Mathematical Society (1996).

\bibitem{Blau:1996ct}
M.~Blau and G.~Thompson, \emph{{Localization and diagonalization: A review of functional integral techniques for low dimensional gauge theories and topological field theories}},  in \emph{{NATO Advanced Study Institute on Functional Integration: Basics and Applications}}, pp.~363--410, 9, 1996.

\bibitem{Jockers:2012dk}
H.~Jockers, V.~Kumar, J.M.~Lapan, D.R.~Morrison and M.~Romo, \emph{{Two-Sphere Partition Functions and Gromov-Witten Invariants}}, \href{https://doi.org/10.1007/s00220-013-1874-z}{\emph{Commun. Math. Phys.} {\bfseries 325} (2014) 1139} [\href{https://arxiv.org/abs/1208.6244}{{\ttfamily 1208.6244}}].

\bibitem{Griguolo:1998kq}
L.~Griguolo, \emph{{The Instanton contributions to Yang-Mills theory on the torus: Localization, Wilson loops and the perturbative expansion}}, \href{https://doi.org/10.1016/S0550-3213(99)00089-9}{\emph{Nucl. Phys. B} {\bfseries 547} (1999) 375} [\href{https://arxiv.org/abs/hep-th/9811050}{{\ttfamily hep-th/9811050}}].

\bibitem{Benini:2013xpa}
F.~Benini, R.~Eager, K.~Hori and Y.~Tachikawa, \emph{{Elliptic Genera of 2d ${\mathcal{N}}$ = 2 Gauge Theories}}, \href{https://doi.org/10.1007/s00220-014-2210-y}{\emph{Commun. Math. Phys.} {\bfseries 333} (2015) 1241} [\href{https://arxiv.org/abs/1308.4896}{{\ttfamily 1308.4896}}].

\bibitem{Griguolo:2024ecw}
L.~Griguolo, R.~Panerai, J.~Papalini, D.~Seminara and I.~Yaakov, \emph{{Localization and resummation of unstable instantons in 2d Yang-Mills}}, \href{https://doi.org/10.1007/JHEP06(2024)188}{\emph{JHEP} {\bfseries 06} (2024) 188} [\href{https://arxiv.org/abs/2403.00053}{{\ttfamily 2403.00053}}].

\end{thebibliography}

\end{document}